\documentclass[times]{rncauth}

\usepackage{moreverb}

\usepackage[dvips,colorlinks,bookmarksopen,bookmarksnumbered,citecolor=red,urlcolor=red]{hyperref}

\newcommand\BibTeX{{\rmfamily B\kern-.05em \textsc{i\kern-.025em b}\kern-.08em
T\kern-.1667em\lower.7ex\hbox{E}\kern-.125emX}}

\def\volumeyear{2010}

\usepackage{subfigure}
\usepackage{multirow}
\graphicspath{{fig/}}

\newcommand{\be}{\ensuremath \mathbf{e}}
\newtheorem{thm}{Theorem}[section]
\newtheorem{cor}{Corllary}[section]

\newtheorem{lem}{Lemma}[section]

\newtheorem{defn}{Definition}[section]
\newtheorem{rem}{Remark}[section]
\newtheorem{assum}{Assumption}[section]
\newtheorem{prob}{Problem}[section]

\begin{document}

\runningheads{H.~ HUANG, C. YU and Q. Wu }{Control of Multi-Agent Formations with Only
Shape Constraints}

\title{Control of Multi-Agent Formations with Only Shape Constraints
\footnotemark[2]}

\author{H.~Huang\footnotemark[3], C. Yu\corrauth and Q. Wu\footnotemark[3]}

\corraddr{Research School of Information Sciences and Engineering, The Australian National University, Canberra ACT 2600, Australia (Brad.yu@anu.edu.au)}

\begin{abstract}This paper considers a novel problem of how to
choose an appropriate geometry for a group of agents with only shape
constraints but with a flexible scale. Instead of assigning the
formation system with a specific geometry, here the only requirement
on the desired geometry is a shape without any location, rotation
and, most importantly, scale constraints. Optimal rigid
transformation between two different geometries is discussed with
especial focus on the scaling operation, and the cooperative
performance of the system is evaluated by what we call the geometries
 \emph{degrees of similarity} (DOS) with respect to the desired shape during the entire convergence process. The design  of the
scale when measuring the DOS is discussed from constant value and time-varying function
perspectives respectively. Fixed structured nonlinear  control laws
that are functions on the scale are developed to guarantee the
exponential convergence of the system to the assigned shape. Our
research is originated from a three-agent formation system and is
further extended to multiple ($n>3$) agents by defining a
\emph{triangular complement graph}. Simulations demonstrate that
formation system with the time-varying scale function outperforms
the one with an arbitrary constant scale, and the relationship
between underlying topology and the system performance is further
discussed based on the simulation observations. Moveover, the
control scheme is applied to bearing-only sensor-target localization
to show its application potentials.
\end{abstract}


\maketitle

\footnotetext[2]{This work was completed during H. Huang's visit at
The Australian National University which was supported by the
University Fund of Beijing Institute of Technology. This work is
also supported by the National Natural Science Foundation of China
under grant 61074031. C. Yu is supported by the Australian Research
Council through Queen Elizabeth II Fellowship under
DP-110100538.}
\footnotetext[3]{School of Automation, Beijing Institute of Technology, Beijing 100081, China (huanghuang,qinghew@bit.edu.cn)}

\vspace{-6pt}


%


\section{Introduction}
\vspace{-2pt}
 A group of agents working together in formation is
seen in various field applications including, for example,
spacecrafts exploring the deep space, underwater vehicles mapping
out oceanbed and unmanned aerial vehicles detecting an unknown
territory. Studies concerning this subject focus primarily on the
stability of the formation systems where there are three  well-known
methodologies, namely distance-based formation control
laws\cite{Anderson07threeForm,Dimarogonas08distanceFormation},
position-based control
laws\cite{Yu09minimalPersistent,Krick09infinitesimallyRigid,Ren08distributed},
and very recently, angle-based
algorithms\cite{Moshtagh09visionFormationBearing,Basiri10angleonly}.
Meanwhile, it is well known that the stability and the performance
of the formation systems depend highly on the underlying
communication topologies. This fact leads to the exploration of
information flow laws among
agents\cite{Fax04,Olfati04consensusSwitching} and  the communication
topology properties of the formation
systems\cite{Anderson08rigidMagazine,Olfati02rigidity}. Consequently
there are control algorithms that aim to deal with various network
dynamics\cite{Olfati04consensusSwitching,Luke09formationDelay,Secchi08formationDelay}.

Current research on formation control relies on the assumption that
the desired geometry is specified, fixed, and known a
priori either in a global coordinate or a local one. Inspired by the
fact that V-shaped formation provides birds with more aerodynamical and visual
advantages than other types of flight in
flocks\cite{Nature01Vformation}, it is reasonable to conjecture that
the geometry formed by the group of agents is closely related to the
behaviors of the formation system. Thus with all those abundant
results under the determined assumptions, we are now ready to move
on to the next stage, that is, to consider situations where the only
concern of the final state is the shape of the system rather than some specific geometry.
In other words, this research considers the scenario where the
requirement on the desired formation is its geometrical information
without any constraints on its location, rotation and scale.

This particular problem arises in angle-based emitter-target
localization  that emerged in
1950s\cite{Bishop09bearingOnlyLocalization}. It were proved in
\cite{Bishop10sensor}, \cite{Dogancay08AOA} and
\cite{Zhang05contour} that there were a set of geometries such as
equilateral triangles in which sensors could provide  the most
accurate estimations of the target position. When the distances
between sensors to the
 target are  sufficiently larger than the ones between sensors, the shape formed by
 the group of agents is then the major concern instead of a specific geometry. It can be further
verified that the geometry with a higher degree of similarity (DOS) (please refer to Section 3 for its definition) to the optimal shape can provide better target location estimation. Thus \emph{in order to improve the accuracy of measurements on the target before sensors reaching the desired shape so as to allow more time for strategy making, a formation  system that maintains a high DOS to the optimal shape during convergence is more preferable than simply focusing on the final shape.}

%
%

However, despite of the potential advantages brought by the
geometrical resemblance, optimal formation control  with only shape
constraints has not yet received enough attentions. To our best
knowledge, there are only a few literatures that are \emph{related}
to this topic. One was reported in
\cite{Spletzer05optimalPositioning} where the authors focused on
finding the optimal scale of the prescribed shape for a set of
robots such that the total distance they traveled was minimized.
However, robot control during the process was ignored and thus the
group of robots did not exhibit any  cooperative behaviors. In
another work\cite{Bhatt09geometric} on optimal relative formation
control of three wheeled robots, the cost function was the kinematic
energy sum of the three robots. By choosing a robot as the rotation
center, the optimal relative positions of robots were designed.
Shape control irrespective of any objectives is seen in a small
number of literatures. By modeling each agent as a double
integrator, reference \cite{Pais09formationShape} considered shape control on
the projected orthogonal reference frame, and very recently,
reference \cite{Basiri10angleonly} proposed a distributed algorithm that
stabilized three agents in the assigned shape. However, those research ignored the performance during the convergence process and thus are not  favorable to tasks such as sensor-target localization.

To this end, this paper aims to explore the novel topic of formation
control under only shape constraints, and exploit the design of
scales that provide the optimal rigid transformation(rotation, translation and scaling operations) between two different geometries. The objective is to minimize  a cost function that integrates the difference between the geometry and the desired shape over the entitle convergence process.
Although there are quite a few discussions on the optimal formation
control, the objectives are normally  fuel
consumption\cite{Kim09pathplanning}, traveling length and
time\cite{Tillerson02}. Fixed structured nonlinear control laws on the
edge state and the pending scale are proposed. When choosing the
appropriate scale of the final geometry so as to minimize the cost
value, two situations are considered one of which is to determine an
optimal value a priori and set it constant during the entire
process. Another more intelligent method is to design a function
named the scale function that adjusts the scale online.

Here is an outline of the paper: Section II contains the notations
and definitions that are used in the paper. A nonlinear control law,
which is the starting point of this research, is introduced in this
section as well. In Section III, the difference between two geometries is defined and
further the cost function is proposed which represents the geometrical performance during formation attainment.
Design of a constant  scale is discussed in Section IV, and in Section V a scale function is  proposed that  further reduces the minimum from the
previous section. By defining a \emph{triangular complement graph}
of the minimally rigid graph, the algorithm applies to multiple
$(n>3)$ agents case as well. In order to broaden the application of
the nonlinear algorithm to situations with determined final
geometries, controllability of the system is discussed by adding a
positive weight factor to the nonlinear control law. Moreover,
applications of the proposed control schema in sensor-target
localization are presented in Section VI. Section VII presents
extensive experimental simulations that verify the effectiveness of
the algorithm and show its application potentials. Conclusions are
made in Section VIII.

\vspace{-6pt}

\section{Preliminary}
\vspace{-2pt}
 Positive real numbers are denoted by $\mathbb{R}^+$
and a matrix of size $m\times n$ is denoted by $\mathbb{R}^{m\times
n}$ where when $n=1$, it is always abbreviated by $\mathbb{R}^m$.
The (block) diagonal matrix of vector $\mathbf{v}$ is denoted by
$\text{diag}(v_i)$ with (vector) $v_i$ being the $i$th (block)
diagonal entry. Sometimes we may also use
$\text{diag}(v_1,\ldots,v_n)$ instead.

  For an undirected graph $G=(V,E)$, the numbers of edges and vertices are
  denoted by $|E|$ and $|V|$ respectively.
Vertex $i$ is the neighbor of vertex $j$, denoted by $i\sim j$, if
they are connected by an edge in $E$. Edge $i$ is the neighbor of
edge $j$ if they share a vertex. The neighbor sets of vertex $i$ and
edge $i$ are denoted by $\mathcal{N}_v(i)$ and $\mathcal{N}_e(i)$
respectively. The minimal distance between vertex $i$ and vertex $j$
is the minimal number of  edges in $E$ that are needed to have them
connected.

Throughout the paper, we focus on 2D formations on a plane, and we
do not consider a system where agents are collinear or coincided,
that is
\begin{assum}\label{assumption_generic}
The initial positions and the desired positions of agents in a
formation system are non-collinear.
\end{assum}

For graph $G$, if we label the nodes in $V$ from 1 to $|V|$ and the
edges in $E$ from 1 to $|E|$, the \emph{representation} of $G$ is a
vector $\mathbf{z}$ whose $i$th entry $z_i\in \mathbb{R}^2$ is the
position of the $i$th node. Given $G$, if we assign an arbitrary
direction on each edge, then the  representation $\mathbf{z}$
uniquely determines an edge vector $\mathbf{e}$ where $e_i\in
\mathbb{R}^2$ corresponds to the $i$th directed edge in $E$, and
$\mathbf{e}$ is called a \emph{relative representation} or a
\emph{geometry} of $G$. Note that in this paper, a geometry is
associated with a graph, which is a more strict definition than that
of a polygon. Given a graph $G$, the distance between two
representations $\mathbf{z}_1$ and $\mathbf{z}_2$ is
$d(\mathbf{z}_1,\mathbf{z}_2)=\max_{i\in V}\|{z_1}_i-{z_2}_i\|$. The
representation $\mathbf{z}$ is further called a \emph{realization}
of $G$ if all the distances between neighboring vertexes in $G$ are compatible, and consequentially $\mathbf{e}$ is the \emph{relative
realization} of $G$.
 All possible realizations of $G$ form a vector space denoted by
$\mathbb{Z}$, and consequently a link/edge space $\mathbb{E}$.  A
\emph{formation} system is a group of agents that communicate over
$G$ and whose deployment is depicted by a realization of $G$.

A realization $\mathbf{z}$ of $G$ in two dimensions is rigid if
there exists $\varepsilon>0$ such that for all realizations
$\mathbf{z}'\in\mathbb{Z}$ that satisfy $\|z'_i-z'_j\|=\|z_i-z_j\|,
\forall i\sim j$ and $d(\mathbf{z},\mathbf{z}')<\varepsilon$, there
holds $\|z'_i-z'_j\|=\|z_i-z_j\|,\forall i,j\in V$. A rigid graph is
further minimally rigid if no single edge can be removed without
losing rigidity. Finally, a globally rigid graph $G$ is a graph
where $\|z'_i-z'_j\|=\|z_i-z_j\|,\forall i,j\in V$ holds for an
arbitrary $\varepsilon>0$. More rigorous definitions could be found,
for example, in
\cite{Yu09minimalPersistent,Anderson08rigidMagazine}.

\begin{figure}
  \subfigure[Similar geometries]{\label{fig:sim_geometry}
\begin{minipage}[b]{0.31\linewidth}
\centering
\includegraphics[scale=0.6]{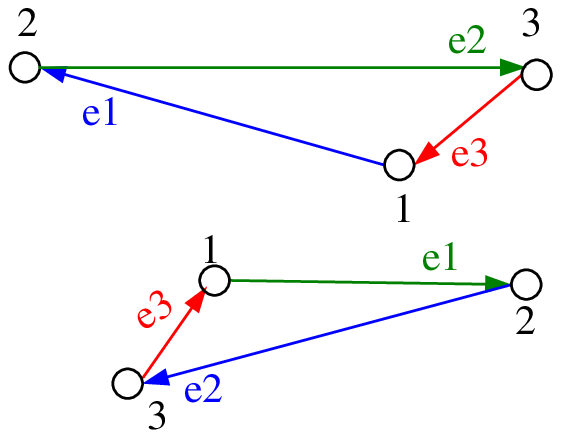}
\end{minipage}}
  \subfigure[Dissimilar geometries]{\label{fig:nonsim_geometry}
\begin{minipage}[b]{0.31\linewidth}
\centering
\includegraphics[scale=0.6]{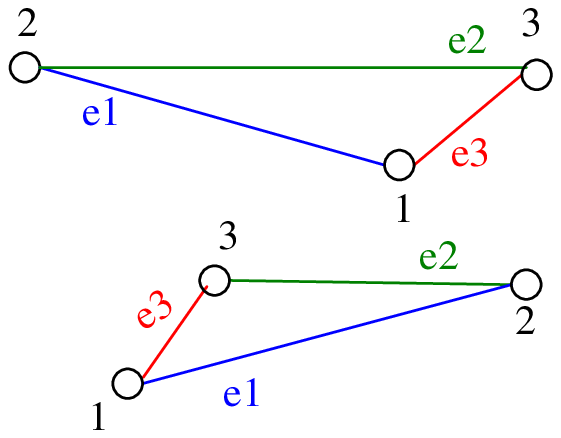}
\end{minipage}}
  \subfigure[Dissimilar geometries]{\label{fig:nonsim_geometry2}
\begin{minipage}[b]{0.31\linewidth}
\centering
\includegraphics[scale=0.6]{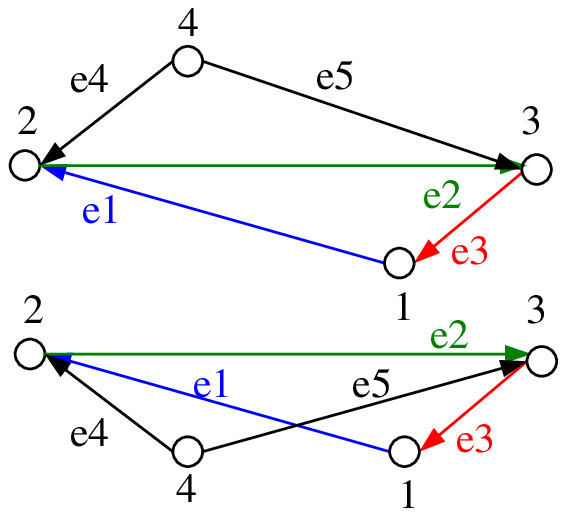}
\end{minipage}}
\caption{Geometries with ordered vertexes}
\end{figure}

Given a graph $G$ and two realizations  $\mathbf{z}_d$ and
$\mathbf{z}$, $\mathbf{z}$ is \emph{similar} to $\mathbf{z}_d$
if
\begin{equation}
   \frac{\|z_l-z_m\|}{\|z_1-z_2\|}=\frac{\|z_{d_l}-z_{d_m}\|}{\|z_{d_1}-z_{d_2}\|},
   \forall l,m\in V
\end{equation}
and consequently, $\mathbf{e}$  and $\mathbf{e}_d$ are \emph{similar}
geometries. Without lose of generality, assume node 1 in both
realizations is deployed at the origin. Then it is necessary that
\begin{equation}\label{eq:similar}
\mathbf{z}_{d_i}=k O(\theta)\mathbf{z}_i
\end{equation}
where $k\in\mathbb{R}$ and $O(\theta)$ is a rotation transformation
through angle $\theta$. Fig. \ref{fig:sim_geometry} and Fig.
\ref{fig:nonsim_geometry} give examples showing similar and
dissimilar geometries. Although the two triangles in Fig.
\ref{fig:nonsim_geometry} are similar ones in the usual sense of
plane geometry, in our case this is no longer true due to the mismatching of indexed nodes.

\begin{figure}
  \centering
  \includegraphics[width=6.7cm,height=1cm]{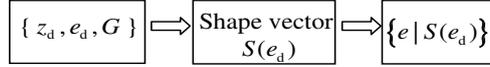}\\
  \caption{The relationship between the shape and the geometry}\label{fig:expla_shape_geome}
\end{figure}

Given a relative realization $\be_d$ of $G$, a shape vector
$S(\mathbf{e}_d)$ generated from $\mathbf{e}_d$ is determined by
$$S(\mathbf{e}_d)=\begin{bmatrix}s_1&s_2&\cdots&s_l\end{bmatrix}^T\in\mathbb{R}^l,l=|E|,s_i\in
\mathbb{R}^+$$ where $s_i=\|e_{d_i}\|$. In most of cases, we will
use $S$ for abbreviation. The set $\{\mathbf{e}|S\}$ consists of all
similar geometries to $S$ over the same graph. Fig.
\ref{fig:expla_shape_geome} gives an explanation on the
relationships between those three notations. Elements are uniquely
determined along the arrows and the reverse is not true. In fact,
given a shape vector $S$, it uniquely determines a shape or a
geometry $\mathbf{e}_d$ if and only if the underlying graph is a
globally rigid one\cite{Anderson08rigidMagazine}.

Due to the nature of the algorithms introduced in this paper and for
convenience, the scale of a geometry in $\{\mathbf{e}|S\}$ with
respect to $S$ is
$$s=\frac{\|e_1\|^2}{2s_1^2}$$
which is  identical to all edges.

The incidence matrix is frequently used to describe the structure of
a directed graph. For an  undirected graph, by assigning each edge
an arbitrary direction, we have the oriented incidence matrix as
seen, for example, in \cite{Diestel10}.
\begin{defn}[Oriented Incidence Matrix]
For an undirected graph $G=(V,E)$, each edge is assigned with a
random direction to generate a directed graph $\bar{G}=(V,\bar{E})$.
The oriented incidence matrix of $G$, denoted by $H(G)$, is a matrix
of dimension $|E|\times |V|$ whose $i$th row corresponds to the
$i$th edge in $\bar{E}$ with entries
\begin{equation*}
  h_{ij}=\begin{cases}
    1&, \bar{e}_i \text{ sinks at } j \\
    -1&, \bar{e}_i \text{ leaves }j\\
    0&, { others }
  \end{cases}
\end{equation*}
\end{defn}
\begin{lem}\label{lemma:rankofH}
  The oriented incidence matrix $H(G)$ has rank $n-1$ when $G$ is connected.
\end{lem}

  By introducing the expended incidence matrix $\hat{H}=H\otimes
  I_2$, where $\otimes$ is the Kronecker product, the edge space $\mathbb{E}$
   can then be considered as a
  codomain of $\hat{H}$ from the vector space $\mathbb{Z}$:
  $$\mathbf{e}=\hat{H}\mathbf{z}$$ where $\mathbf{z}\in \mathbb{Z}$ and $\mathbf{e}\in
  \mathbb{E}$. Thus we have $\mathbf{e}\in \text{Im} \hat{H}$.

Further we define a rigidity function\cite{Yu09minimalPersistent}
(also known as the edge function\cite{Dorfler09cooperativeBehavior})
\begin{equation}\label{eq:rigidityFun}
\bar{\mathbf{z}}=r(\mathbf{e})=\frac{1}{2}\begin{bmatrix}\|e_1\|^2&\cdots&\|e_m\|^2\end{bmatrix}^T
\end{equation}
and the rigidity matrix is further defined as
$$R(\mathbf{e})=\frac{\partial \bar{\mathbf{z}}}{\partial \mathbf{z}}=\Lambda(\mathbf{e})^T\hat{H}$$ with
$\Lambda(\mathbf{e})=\text{diag}(e_k)$, $e_k\in\mathbb{R}^2$. For a
rigid graph, $\text{rank}R(\mathbf{e})=2|E|-3$.



\begin{lem}[\cite{Dorfler09cooperativeBehavior}]\label{lemma:dorfler}
For a desired relative realization $\mathbf{e}_d$ over $G=(V,E)$ and
the vector $\mathbf{d}\in\mathbb{R}^{|E|}$ with
$d_k=[\frac{1}{2}\|e_{d_k}\|^2]$, by choosing the potential function
\begin{equation}
  V(\mathbf{e})=\sum_{k=1}^{|E|}\frac{1}{8}(\|e_k\|^2-2d_k)^2
\end{equation}
the control law
\begin{equation}\label{eq:rigidoptU}
  \dot{\mathbf{z}}=u=-\hat{H}^T [\partial V(\mathbf{e})/\partial \mathbf{e}]^T
\end{equation}
is an \emph{inverse} optimal solution to the optimization problem
\begin{align}\label{eq:rigidoptJ}
  &\min\bar{J}(\mathbf{e}_0,u)=\frac{1}{2}\int^\infty_0\|R(\mathbf{e})^T[r(\mathbf{e})-\mathbf{d}]\|^2+\|u\|^2d\tau\nonumber\\
&\text{s.t. } \dot{\mathbf{e}}=\hat{H} u,
\mathbf{e}_0=\mathbf{e}(0)\in \text{Im}\hat{H}
\end{align}
and the formation system converges to the largest  invariant set
\begin{equation}\label{eq:invariantset_dor}
\mathcal{I}_e=\{\mathbf{e}\in \text{Im} \hat{H}:
V(\mathbf{e})<V(\mathbf{e}_0),
\|R(\mathbf{e})^T[r(\mathbf{e})-\mathbf{d}]\|=0 \}\end{equation}
\end{lem}
The proof of the lemma is based on the Hamilton-Jacobi-Bellman (HJB)
equation and Lyapunov Theory. For details please refer to
\cite{Dorfler09cooperativeBehavior}.

Algorithm \eqref{eq:rigidoptU} requires that each agent has the
knowledge of the global coordinates of its neighbors. This
requirement is also assumed to be true in our research.

\section{Cost function and cooperative performance}
With the wide range of applications of sensor networks, the geometric characters of  a group of sensors are sometimes crucial to the achievement of tasks, such as bearing only sensor-target localization, shape-constrained formation reconfiguration and obstacle avoidance. Thus it is exigent to propose an  efficient way to distinguish two geometries.

In image processing, especially in pattern matching, the Hausdorff distance is commonly used to evaluate how much do two geometries resemble each other under rigid transformation\cite{Hutten93hausdorff}. The problem of shape matching is to find the optimal transformation that minimizes the distance. Sometimes points on the edges are assigned with different weights indicating their importance to the shape. However, the Hausdorff distance metric is on the basis a minimization function, which is  not favorable as an objective function. Meanwhile, at each  time step, the time complexity to compute the Hausdorff distance  for two points sets of size $p$ and $q$ is $O(pq)$. Thus another metric function is expected that produces intuitively reasonable results and is beneficial to optimization.

\begin{figure}
 \subfigure[When the short edge suffer a small perturbation, the perturbed geometry various a lot w.r.t. the nominal one; When the same perturbation is added to the long edge, the new geometry is quite close to the nominal one]{\label{fig:DOS_lengths}
\begin{minipage}[b]{0.45\linewidth}
\centering
\includegraphics[scale=0.4]{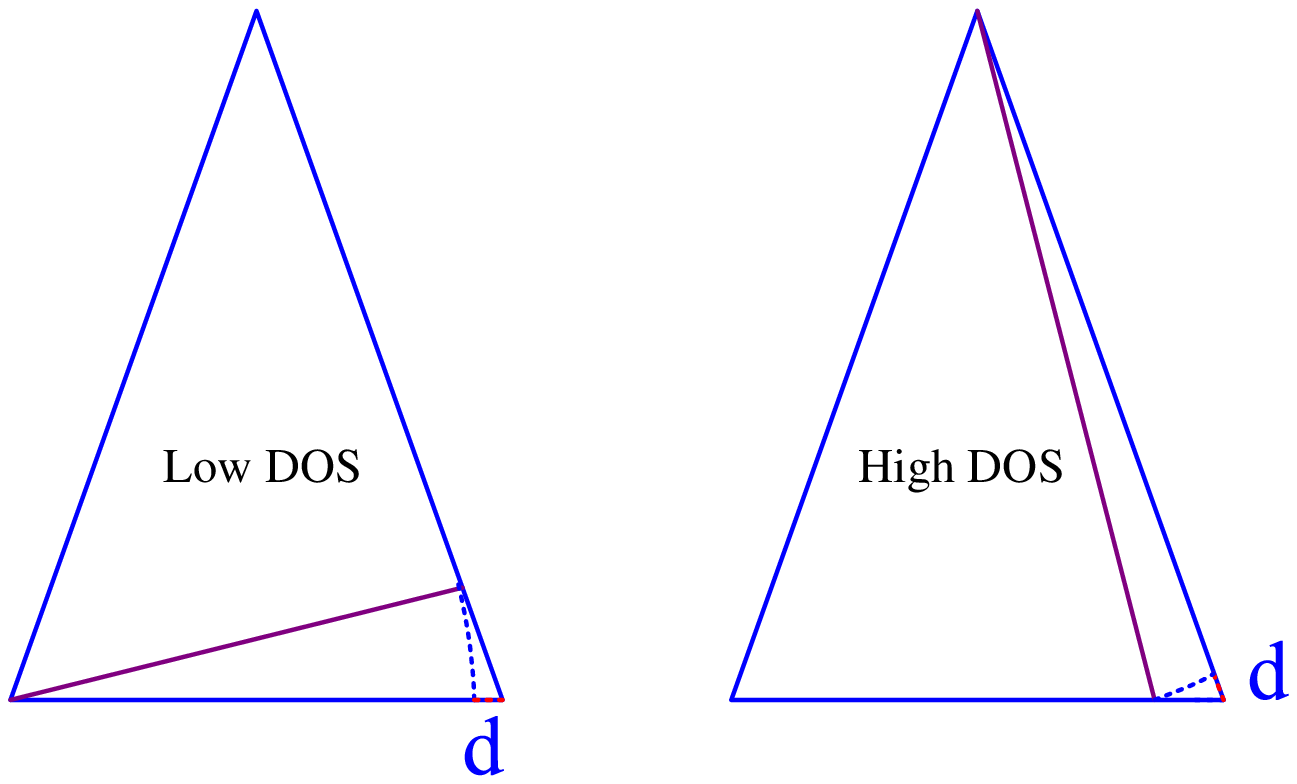}
\end{minipage}}
  \subfigure[When the angle included in the  two edges is large($0~\pi$), perturbations on the two edges will result in small change to shape; When the angle is small($0~\pi$), the shape is greatly affected by the same perturbations]{\label{fig:DOS_angles}
\begin{minipage}[b]{0.45\linewidth}
\centering
\includegraphics[scale=0.4]{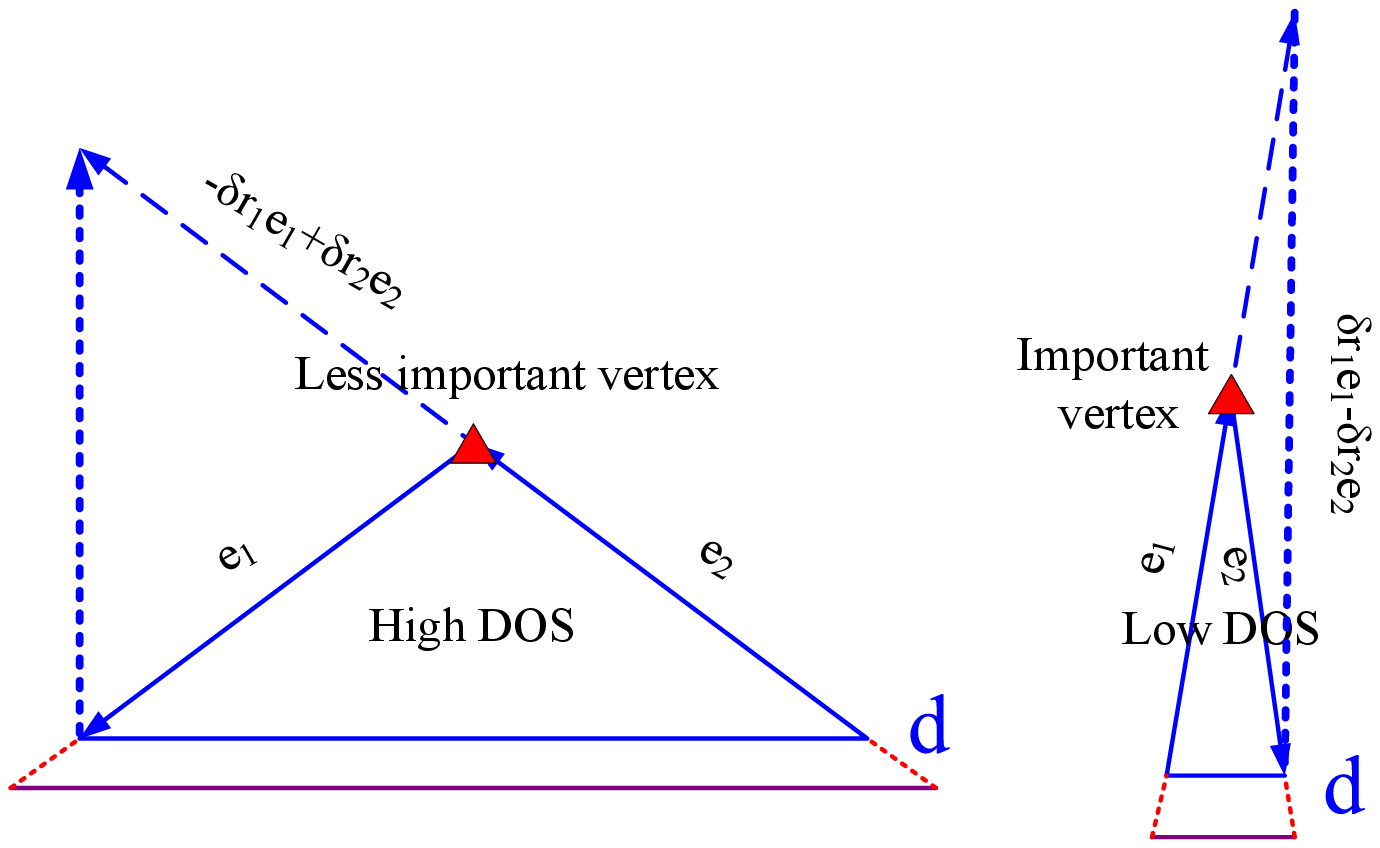}
\end{minipage}}
\caption{The sensitivity  of a geometry}\label{fig:DOS}
\end{figure}

Before we dive into the mathematical formalizations, here are two observations for determining the resemblance of  geometries with respect to the reference/desired shape:
\begin{itemize}
  \item [i] A geometry is less sensitive to perturbations on \emph{long} edges
  \item [ii] A geometry is less sensitive to perturbations on edges with large included angles $(0,\pi)$
\end{itemize}
Fig. \ref{fig:DOS} further explains the above two observations.
It is intuitive that  under the same perturbation, the long edge bring small deformation to the geometry
than that of a short edge, as the two geometries in Fig. \ref{fig:DOS_lengths}.
Meanwhile, apart from the lengths of edges, another profile of a shape is the vertex,
especially those sharp vertexes, i.e., the included angle of the two  edges adjacent at  the vertex is small.

 When two geometries are similar to each other, the scale between them is the ratio of two corresponding edges in the geometries and is identical to all  the pairs of edges. However, for two dissimilar geometries, there is no common ratio for any pairs of edges, thus choosing an appropriate scale is premise to the analysis of geometries resemblance.

According to the above two observations, it is obviously inappropriate to simply use the differences between edges lengths of the two geometries to measure how two geometries are related:
\begin{equation}
J'=\int_0^\infty \|r(\mathbf{e})-r(\be')\|^2 d\tau
\end{equation}
because $J'$ dose not consider observation $i$ or observation $ii$.

We are interested in the resemblance of two geometries, thus under observations $i$ and $ii$, the difference between geometry $\be$ and $\be'$ is
$$\rho_\Theta(\be,\be')=H^T\begin{bmatrix}(r_1(\be) -  \Theta r_1(\be')) e_1\\ \vdots \\ (r_n(\be)- \Theta r_n(\be')) e_n\end{bmatrix}$$
As one can tell, an appropriate scale $\Theta$ is key to the evaluation. In our research, the geometry is depicted by $2|V|-3$ directed edges
rather than the coordinates of the vertexes, thus the translation
and rotation operations are not our concerns.
 The element $(r_i(\be) -  \Theta r_i(\be')) e_i:=\delta r_i e_i$ indicates that
 each edge in one geometry is scaled by the edge difference between the two geometries.
 This is designed to take care of observation $i$.b
Inspired by the weighted Hausdorff distance, we introduce the transformation $H^T$ which allows two neighboring edges communicate and  consequently, as shown in Fig. \ref{fig:DOS_angles}, evolve into the edges in dot lines, the lengths of which are positive monotony to the sharpness of the vertexes.

Given a geometry  $\be$  and a shape $S(\be')$ , the degree of similarity (DOS) of $\be$ with respect to $S$ is measured by
\begin{equation}\label{eq:distance}
dos_\Theta(\be,S(\be'))=\|\rho_\Theta(\be, \be')\|_2^{-2}
\end{equation}
A geometry with a higher DOS to $S$ is said to be more resemble to $S$. When the geometry is similar to $S$, $dos_\Theta(\be,\be')\rightarrow \infty$.

As mentioned before, by taking care of the geometries DOS during the entire convergence, sensors can provide reliable measurement of the target in an early stage.  Thus in this research, the \emph{geometrical performance of a formation system is the integral of the DOS with respect to the desired shape during convergence}:
\begin{equation}\label{eq:myJ0}
  J(\mathbf{e}_0,u,\Theta)=\int^\infty_0 dos^{-1}_\Theta(\be,S(\be'))d\tau=\int^\infty_0 \|R(\mathbf{e})^T[r(\mathbf{e})-\Theta\bar{S}]\|^2 d\tau
\end{equation}
with $\bar{S}=[\bar{s}_i]$ and $\bar{s}_i=s_i^2$. We use $S^2$
instead of $S$ due to the composition of $r(\mathbf{e})$ in
\eqref{eq:rigidityFun}. Apart from the initial state $\mathbf{e}_0$
and the control law $u$, the value of $J$ is also a function on the scale $\Theta$.

\begin{figure}
 \subfigure[Performance]{\label{fig:cooperative_good_bad}
\begin{minipage}[b]{0.43\linewidth}
\centering
\includegraphics[scale=0.6]{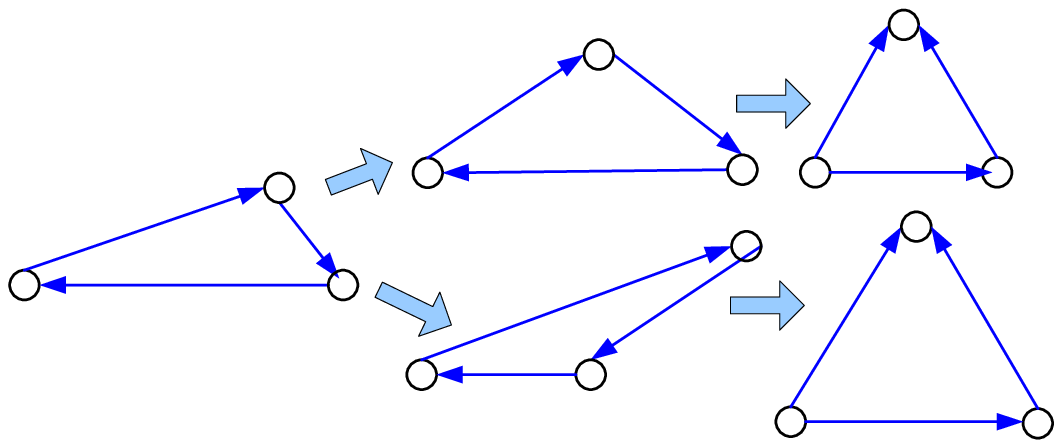}
\end{minipage}}
  \subfigure[Triangle]{\label{fig:three_topo}
\begin{minipage}[b]{0.24\linewidth}
\centering
\includegraphics[scale=0.5]{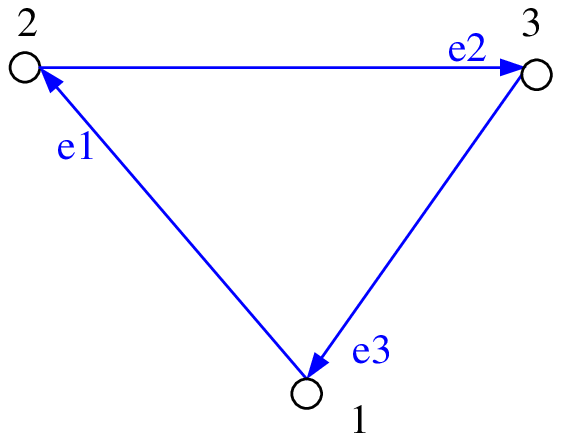}
\end{minipage}}
  \subfigure[New space]{\label{fig:three_topo2}
\begin{minipage}[b]{0.23\linewidth}
\centering
\includegraphics[scale=0.5]{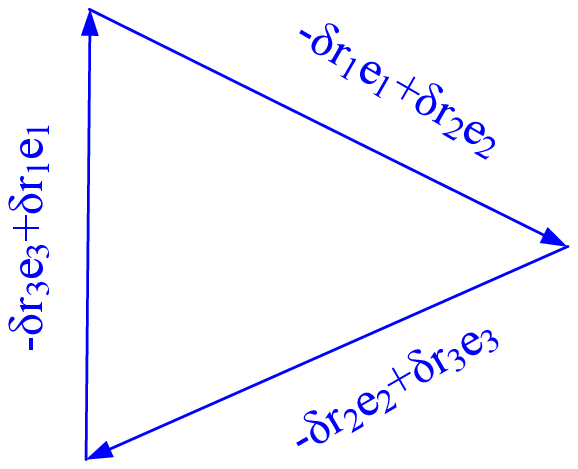}
\end{minipage}}
\caption{Triangular geometry of three agents. For the definitions of DOS, please refer to \eqref{eq:distance}}\label{fig:unknown}
\end{figure}

For example  we consider three agents in a formation as shown in Fig.
\ref{fig:three_topo}. If we set $\delta
r_i=r_i-\Theta\bar{s}_i$ and denote
$L=\|R(\mathbf{e})^T[r(\mathbf{e})-\Theta\bar{S}]\|^2$, it yields
\begin{align}
   L=&\|\hat{H}^T\begin{bmatrix}
    \delta r_1 e_1 &\cdots & \delta r_n e_n
  \end{bmatrix}^T\|^2\nonumber\\
  =&\|-\delta r_1 e_1+\delta r_2 e_2\|^2+\|-\delta r_2 e_2+\delta r_3
e_3\|^2+\|\delta r_1 e_1-\delta r_3 e_3\|^2
\end{align}
The value of $L$ is the sum of  lengths of the three new vectors in Fig. \ref{fig:three_topo2}.

\begin{rem}
A formation system with a smaller cost value \eqref{eq:myJ0} is considered to
exhibit better geometrical/cooperative performance during convergence.
\end{rem}

By minimizing the cost function \eqref{eq:myJ0}, the two situations as, for example, the ones in Fig.
\ref{fig:cooperative_good_bad}, are expected to be distinguished. The upper case has better geometric
performance as the geometries during convergence have consistently high DOS
 with respect to the desired shape. For the bottom
situation, although the three agents attain the desired geometry, it
differs a lot from the desired shape during the process. In sensor
networks for localization, formation system in the upper case may
allow sensors to provide some rough estimate of the target before
attaining the formation. We will show this later through
experimental results in Section \ref{sec:exp}.

 When the
underlying graph of the formation system
 is a minimally rigid graph, the convergence of $J$ is equivalent
to $r(\mathbf{e})\rightarrow \Theta \bar{S}$. Two forms of $\Theta$
are considered in this paper, namely $\Theta$ being constant:
$\Theta=s_c\in\mathbb{R}^+$, and $\Theta$ being a time-varying
function on $\mathbf{e}$: $\Theta=\tilde{s}(\mathbf{e})\in
C(\mathbb{E})$, where $C$ is the set of continuous mappings. The cost
functions are then written, respectively, as:
\begin{equation}\label{eq:jc}
  J_c(\mathbf{e}_0,u,s_c)=\int^\infty_0 \|R(\mathbf{e})^T[r(\mathbf{e})-s_c\bar{S}]\|^2 d\tau
  \end{equation}
  and
\begin{equation}\label{eq:myJ}
  J_v(\mathbf{e}_0,u,\tilde{s}(\be))=\int^\infty_0 \|R(\mathbf{e})^T[r(\mathbf{e})-\tilde{s}(\mathbf{e})\bar{S}]\|^2 d\tau
\end{equation}

The main concern in this research is the determination of $s_c$ and
$\tilde{s}(\mathbf{e})$  such that the cost functions $J_c$ and $J_v$
are minimized. Note that $J_c$ and $J_v$ are not just quadratic
expressions but a function on $\mathbf{e}_0$, $u$ and $s_c$ or
$\tilde{s}(\mathbf{e})$, thus minimizing the values of $J_c$ and
$J_v$ is not trivial. Inspired by the control law in Lemma
\ref{lemma:dorfler}, in this research, we will fix the structure of
the control law and focus on the design of the optimal scale
function.
\begin{prob}\label{prob:general}
  Consider a formation system where each agent is modeled by
\begin{equation}\label{eq:basic_sys}
  \dot{\mathbf{z}}=u
\end{equation}
and the underlying graph $G$ is a minimally rigid graph.
\begin{itemize}
\item Let $\Theta=s_c\in \mathbb{R}^+$, and consider the following control law on   $s_c$ and $\mathbf{e}$:
\begin{equation}\label{eq:myde_eqS}
u(s_c,\mathbf{e})=-R^T(\mathbf{e})[r(\mathbf{e})-s_c\bar{S}].
\end{equation}
Find the optimal value of $s_c$ such that $\be_f\in\{\be|S\}$ where
$\be_f$ is the stable realization and
\begin{equation}\label{eq:jc0}
J_c^*(\mathbf{e}_0,u(s_c,\mathbf{e}))=\min\limits _{s_c\in
\mathbb{R}^+} J_c(\mathbf{e}_0,u(s_c,\mathbf{e}),s_c)
\end{equation}
\item Let $\Theta=\tilde{s}(\mathbf{e})$ where $\tilde{s}\in C(\mathbb{E})$, and consider the following control
law on  $\tilde{s}(\mathbf{e})$ and $\mathbf{e}$:
\begin{equation}\label{eq:u_varying}
  u(\tilde{s}(\mathbf{e}),\mathbf{e})=-R(\mathbf{e})^T M(\mathbf{e})^T
  [r(\mathbf{e})-\tilde{s}(\mathbf{e})\bar{S}].
\end{equation}
Find the control gain $M(\mathbf{e})\in\mathbb{R}^{(2n-3)\times
(2n-3)}$ and the scale function $\tilde{s}(\mathbf{e})$ such that
$\be_f\in\{\be|S\}$ and
\begin{equation}\label{eq:jv0}
J_v^*(\mathbf{e}_0,u(\tilde{s}(\mathbf{e}),\mathbf{e}))=\min\limits
_{\tilde{s}\in C(\mathbb{E})}
J_v(\mathbf{e}_0,u(\tilde{s}(\mathbf{e}),\mathbf{e}),\tilde{s}(\mathbf{e}))
\end{equation}
\end{itemize}
\end{prob}

When $\Theta=s_c$, it is actually a matching of two fixed geometries without considering the scaling operation. When $\Theta=\tilde{s}(\mathbf{e})$, it turns to a matching of a geometry and a  shape, which gives us more flexibility to optimize  the geometry's DOS.


In the remaining context of the paper, we will discuss the
optimization problem \eqref{eq:jc0} and \eqref{eq:jv0} respectively.

\section{Optimal formations under time-invariant shape
scale}\label{sec:inva} System with a constant scale $s_c$ during the
entire converging process is simple but typical. Instead of
considering a determined final geometry at a randomly selected scale
as discussed in Lemma \ref{lemma:dorfler}, here given the initial
relative realization $\be_0$, we focus on finding an optimal $s_c$
 such that the formation system has the best cooperative performance.

When $\Theta=s_c$ and under the control law \eqref{eq:myde_eqS},
\begin{equation}\label{eq:constant_J1}
  J_c(\mathbf{e}_0,u(\be,s_c),s_c)=\sum^n_{i=1}\frac{1}{8}(\|e_i(0)\|^2-2s_c\bar{s}_i)^2
\end{equation}
The partial derivative of $J_c$ with respect to $s_c$ is
\begin{equation}\label{eq:partialOfStaticS}
  \frac{\partial J_c}{\partial
  s_c}=\sum_{i=1}^n\frac{1}{2}(\|e_i\|^2-2s_c\bar{s}_i)\bar{s}_i
  \end{equation}
  By letting \eqref{eq:partialOfStaticS} equal to zero, it immediately yields
  that the optimal  solution $s^*_c$ to problem \eqref{eq:jc0} is
  \begin{equation}
    \sum_{i=1}^n(\|e_i(0)\|^2-2s_c^*\bar{s}_i)\bar{s}_i=0
  \end{equation}
  Under Assumption \ref{assumption_generic}, the analytical expression of the optimal scale is
  \begin{equation}\label{eq:optStaticS}
    s^*_c=\frac{\sum_{i=1}^n\|e_i(0)\|^2\bar{s}_i}{2\sum_{i=1}^n \bar{s}_i^2}
  \end{equation}
  and it is obvious that $s^*_c\in\mathbb{R}^+$.


Rigidity corresponds to a unique realization of the graph only if
the domain is small enough. For a globally unique realization, we
have the idea of global rigidity\cite{Anderson08rigidMagazine} which
tells that once the distances between any of the two neighboring
agents are known, all the other distances for non-neighbors are
uniquely determined. However, globally rigid graph is rigorous and
there is no polynomial time algorithm to determine weather a graph
is a globally rigid one or not. Meanwhile, its rigidity matrix is
not a full rank matrix, which may brought some unexpected
equilibriums in the rigidity matrix-based control laws. Thus we
would like to focus ourselves on minimally rigid graph whose
rigidity matrix is exactly of rank $2|V|-3$.

\begin{thm}
The formation system \eqref{eq:basic_sys} over a minimally rigid
graph converges to the largest invariant set
\begin{equation}
\mathcal{I}_e=\{\mathbf{e}\in \text{Im} \hat{H}:
r(\mathbf{e})-s^*_c\bar{S}=0,\mathbf{z}_0\in\Omega_\sigma\}
\end{equation}
where
\begin{align}\label{eq:omega}
\Omega_\sigma(S):=&\{\mathbf{e}_0\in \mathbb{E}:
\sum_{i=1}^n(r_i(\mathbf{e}_0)-r_i(\mathbf{e}))^2+\nonumber\\
&\sum_{l\not\sim
m}(\frac{1}{2}\|z_{0_l}-z_{0_m}\|^2-\frac{1}{2}\|z_l-z_m\|^2)^2<\sigma,\mathbf{e}\in\{\mathbf{e}|S\}\}
\end{align}
for a sufficiently small $\sigma$ if the control law  is the one in
\eqref{eq:myde_eqS}. Furthermore, $J^*_c$ is obtained when
$s_c=s^*_c$ with $s^*_c$ given in \eqref{eq:optStaticS}.
\end{thm}
\begin{proof}
  Similar to the proof of Lemma \ref{lemma:dorfler},
  the control law \eqref{eq:myde_eqS} for an arbitrary  $s_c\in\mathbb{R}^+$
   is a stabilization control law that guarantees the convergence to
   $\mathcal{I}_e$.
The derivation of $s^*_c$ indicates that \eqref{eq:optStaticS} is
the optimal solution to \eqref{eq:jc0}, which finishes the proof.
\end{proof}
\begin{rem}\label{rem:initial}
 When the underlying graph $G$ of the  shape is a minimally rigid
one, a shape $S$ dose not determines a geometry uniquely, and
consequently $r(\mathbf{e}_f)=s^*_c\bar{S}$ dose not ensures
$\mathbf{e}_f\in\{\mathbf{e}|S\}$. However, when the initial
geometry $\mathbf{e}_0$ of the formation system is restrained within
a sufficiently small neighborhood $\Omega_\sigma(S)$ of a candidate
geometry, $\dot{V}(\mathbf{e})<0$ guarantees the convergence to the
desired shape.
\end{rem}

The optimal scale $s^*_c$ is, not surprisingly, uniquely determined
by the initial geometry $\mathbf{e}_0$ and the desired shape $S$. It
is the global optimum and is always positive with an upper bound
when the formation is non-collinear.

In a formation system, when a set of agents equipped with sensors is
deployed within a neighborhood of the desired shape specified by the
shape vector $S(G)$, \eqref{eq:optStaticS} helps us to decide the
exact final geometry that requires a small range of transformations
from $\be_0$, and the control law drives the agents approach this
optimal geometry.

\section{Formation control under time-varying scale
function}\label{sec:varying}
As mentioned previously, there are two kinds of strategies to find
an appropriate shape scale one of which  is to determine an optimal
value a priori and set it constant during the entire process, as
discussed in the previous section. Another more intelligent strategy
is to consider the time-varying function $\tilde{s}$ that adjusts
the final scale online. This allows us to evaluate the resemblance of the geometry with respect to the desired shape more objective.

\subsection{Formations with three agents}
 Triangular
formations with three agents interconnected with one another are the
most fundamental pattern in formation systems, and thus are
embraced as the starting point in many literatures, as for example
in
\cite{Yu09minimalPersistent,Huang10optimalFormation,Anderson08UAV}.
Here we will also start with this simple and typical case and then
further extend the results into multiple $(n>3)$ agents systems.

Under the fixed-structured control law  \eqref{eq:u_varying}, the
dynamics of the formation system in the edge space is
\begin{equation}\label{eq:mydes_M}
  \dot{\mathbf{e}}=\hat{H}u=-\hat{H}R(\mathbf{e})^T M(\mathbf{e})^T [r(\mathbf{e})-\tilde{s}(\mathbf{e})\bar{S}]
\end{equation}
The control law $u$ is a function of both $\tilde{s}(\mathbf{e})$
and $\mathbf{e}$, and $\mathbf{e}$ in $\tilde{s}(\mathbf{e})$ is
also determined by $u$, which means $u$ and $\tilde{s}(\mathbf{e})$
are highly coupled. This makes the solution to \eqref{eq:jv0}
becomes complex and the problem may not be solvable  exactly.

Alternatively, we would like to seek for a function
$\tilde{s}(\mathbf{e})$ for the following problem:
\begin{prob}\label{prob:threeOptimal}
Find a trajectory for $\tilde{s}\in C(\mathbb{E})$, which is an
explicit function of edge $\mathbf{e}$, such that
$$J_v(\mathbf{e}_0,u(\tilde{s}(\mathbf{e}),\be),\tilde{s}(\mathbf{e}))<J_c((\mathbf{e}_0),u(s_c,\be),s_c)$$
for all $s_c\in\mathbb{R}^+$.
\end{prob}
Instead of concentrating on the complex optimization problem
\eqref{eq:jv0}, we suggest to design an  adjusting rule for the scale such that the formation system under the control law always exhibits better
cooperative performance than the case with a constant scale.

Consider the equation
\begin{equation}\label{eq:temp1}
  \frac{\partial J_v}{\partial \tilde{s}}\mid_{\tilde{s}=\tilde{s}^*(\mathbf{e})}=0
\end{equation}
which has the equivalent form of
\begin{equation}
  \frac{\partial L}{\partial \tilde{s}}=2[r(\mathbf{e})^T-\tilde{s}(\mathbf{e})\bar{S}^T]
  R(\mathbf{e}) R(\mathbf{e})^T \bar{S}
  \end{equation}
  where $L$ is the loss function of $J_v$.
By using the fact  $$e^T_i e_j=\|e_i\|\|e_j\|cos\theta_{ij}$$ and
the cosine law
$$\|e_i\|\|e_j\|cos(\theta_{ij})=\frac{1}{2}(\|e_i\|^2+\|e_j\|^2-\|e_k\|^2)$$
where $e_i$, $e_j$ and $e_k$ form an triangle, we obtain
\begin{equation}
  \frac{\partial L}{\partial \tilde{s}}=2[r(\mathbf{e})-\tilde{s}(\mathbf{e})\bar{S}]^T \bar{D} r(\mathbf{e})
\end{equation}
where
\begin{equation}\label{eq:D_three}\bar{D}=\begin{bmatrix}
\bar{s}_2+\bar{s}_3-4\bar{s}_1& \bar{s}_2-\bar{s}_3& \bar{s}_3-\bar{s}_2\\
\bar{s}_1-\bar{s}_3 & \bar{s}_1+\bar{s}_3-4\bar{s}_2 & \bar{s}_3-\bar{s}_1\\
\bar{s}_1-\bar{s}_2 &\bar{s}_2-\bar{s}_1
&\bar{s}_1+\bar{s}_2-4\bar{s}_3\end{bmatrix}\end{equation} If we let
$\frac{\partial L}{\partial \tilde{s}}=0$,  a trajectory of
$\tilde{s}(\mathbf{e})$ is then
\begin{equation}\label{eq:mydynS}
  \tilde{s}^*(\mathbf{e})=\frac{r(\mathbf{e})^T \bar{D} r(\mathbf{e})}{\bar{S}^T \bar{D} r(\mathbf{e})}:\triangleq
  \frac{s_N}{s_D}
\end{equation}
and next we prove that the cost function $J_v$ with the scale
 \eqref{eq:mydynS} is consistently smaller than $J_c$.
\begin{thm}\label{thm:JvJc}
  Given the same initial geometry $\be_0$ and the same desired shape $S$, the inequality
  $$J_v(\mathbf{e}_0,u(\tilde{s}^*(\mathbf{e}),\be),\tilde{s}^*(\mathbf{e}))\leq J_c((\mathbf{e}_0),u(s_c,\be),s_c)$$
  with $\tilde{s}^*$ given in \eqref{eq:mydynS}
  always holds true for arbitrary $s_c\in \mathbb{R}^+$. The
  equality is satisfied if and only if $\mathbf{e}_0\in
  \{\mathbf{e}|S\}$.
\end{thm}
\begin{proof}
  Substituting $\tilde{s}^*$ into $J_v$, the difference of $J_v$ and
  $J_c$ is
  \begin{align}
    &J_v-J_c\nonumber\\
    &=\int_0^\infty\left((r-\tilde{s}^*\bar{S})^TRR^T(r-\tilde{s}^*\bar{S})-
(r-s_c\bar{S})^TRR^T(r-s_c\bar{S})\right)d\tau\nonumber\\
           &=\int_0^\infty[{(\tilde{s}^*}^2-s_c^2)\bar{S}^T+2(s_c-\tilde{s}^*)r^T]RR^T\bar{S}d\tau\nonumber\\
           &=\int_0^\infty(\tilde{s}^*-s_c)(\tilde{s}^*\bar{S}^T-r^T+s_c\bar{S}^T-r^T)RR^T\bar{S}d\tau
           \label{eq:JvJc}
  \end{align}
Recall the characters of $\tilde{s}^*$,
\begin{equation}\label{temp}
[r^T-\tilde{s}^*\bar{S}^T]
  R R^T \bar{S}=0
  \end{equation}
Based on the expression of $\tilde{s}^*$ in \eqref{eq:mydynS},
\begin{equation}\label{temp2}
  \tilde{s}^*-s_c=\frac{(r^T-s_c\bar{S}^T)\bar{D}r}{\bar{S}^T\bar{D}r}
\end{equation}
According to \eqref{temp} and \eqref{temp2}, equation
\eqref{eq:JvJc} has the simplified form of
  \begin{equation}
    J_v-J_c=\int_0^\infty -\frac{1}{s_D}\bar{S}^TRR^T(r-s_c\bar{S})(r^T-s_c\bar{S}^T)\bar{D}rd\tau\nonumber\\
  \end{equation}
  The positiveness of $s_D$ is guaranteed according to the equality of
  $$\bar{D}r=RR^T\bar{S}$$ which further
  yields
  \begin{equation}
    J_v-J_c=\int_0^\infty -\frac{1}{s_D}((r^T-s_c\bar{S}^T)\bar{D}r)^2d\tau\leq0,\forall s_c>0
  \end{equation}
  where equality holds if and only if $s_c\equiv\frac{\|e_i\|^2}{2s_i^2},\forall i\in[1,2,3]$,
i.e., $\mathbf{e}_0\in \{\mathbf{e}|S\}$.
\end{proof}

 Even when $s_c=s^*_c$, the conclusion of $J_v$ having a smaller
value still holds true, which indicates that better
cooperative performance is always observed on $J_v$.

\begin{thm}\label{thm:stable}
The triangular formation system \eqref{eq:mydes_M} is exponentially
stable and converges to the invariant set
$\mathcal{I}_\mathbf{e}=\{\mathbf{e}\in\text{Im}\hat{H}:r(\mathbf{e})-\tilde{s}^*_f\bar{S}=0\}$
if
\begin{equation}\label{eq:myM}
  M(\mathbf{e})=s_D^2 I_3-s_D \bar{S} r(\mathbf{e})^T(\bar{D}^T+\bar{D}) +
  s_N
  \bar{S} \bar{S}^T \bar{D}
\end{equation} with parameters $\bar{D}$ and $s_D,s_N$
given in \eqref{eq:D_three} and \eqref{eq:mydynS} respectively, and
$\tilde{s}^*_f$ is the stable value of the scale function
$\tilde{s}(\mathbf{e})$.
\end{thm}
Before giving the proof to Theorem \ref{thm:stable}, we first
propose the following lemma
\begin{lem}\label{lemma:M_nonsigular}
  Matrix $M(\mathbf{e}),\mathbf{e}\in\mathbb{E}$ in \eqref{eq:myM} is a singular matrix
  if and only if $r(\mathbf{e})= k\bar{S}$.
\end{lem}
\begin{proof}
Let $Q$ be the transformation matrix such that
$$Q\bar{S}=\begin{bmatrix} \bar{s}_1&0& 0
\end{bmatrix}^T$$
Left multiplying matrix $M(\mathbf{e})$ by $Q$ yields
\begin{align}
  QM(\mathbf{e})&=s_D^2IQ-s_D\begin{bmatrix}\bar{s}_1\\0\\0\end{bmatrix}
  r(\mathbf{e})^T(\bar{D}^T+\bar{D})+s_N\begin{bmatrix}\bar{s}_1\\0\\ 0\end{bmatrix}\bar{S}^T\bar{D}\nonumber\\
  &=\begin{bmatrix}
    s_D^2-p_1&-p_2&-p_3\\
    -\frac{\bar{s}_2}{\bar{s}_1}s_D^2&s_D^2&0\\
    -\frac{\bar{s}_3}{\bar{s}_1}s_D^2&0&s_D^2
  \end{bmatrix}\label{eq:lemma_temp}
\end{align}
where
$$p_i=s_D\bar{s}_1\sum_{j=1}^{3}r_j(\mathbf{e})(\bar{d}_{ij}+\bar{d}_{ji})-
s_N\bar{s}_1\sum_{j=1}^3\bar{s}_j\bar{d}_{ji}$$ and $\bar{d}_{ij}$
is the elements of matrix $\bar{D}$. Matrix $M(\mathbf{e})$ is of
full rank if and only if the diagonalize matrix
$\text{diag}(m,s^2_D,s^2_D)$ with
$$m=s_D^2-p_1+\sum_{i=2}^3(-\frac{\bar{s}_i}{\bar{s}_1}p_i)$$ consists of only
nonzero diagonal entries, that is $m\neq 0$ or equivalently,
\begin{equation}\label{eq:dp_sd2d1}
  \sum_{i=1}^3(\bar{s}_ip_i)\neq s_D^2\bar{s}_1
\end{equation}
Recall \eqref{eq:lemma_temp}, if we consider $\sum(\bar{s}_ip_i)$ as
the inner product of the two vectors $\mathbf{p}=\begin{bmatrix}p_1
&p_2&p_3\end{bmatrix}^T$ and $\bar{S}$, we obtain
\begin{equation}\label{eq:temp2}
  \sum_{i=1}^3(\bar{s}_ip_i)=s_D \bar{s}_1 r^T(\bar{D}^T+\bar{D}) \bar{S}-s_N \bar{s}_1 \bar{S}^T\bar{D}
  \bar{S}
\end{equation}
Substituting $s_D$ and $s_N$ given in \eqref{eq:mydynS}  into
\eqref{eq:dp_sd2d1} and by some trivial calculations, we conclude
that the necessary and sufficient condition for $M(\mathbf{e})$
being a singular matrix is
\begin{equation}\label{eq:judgeM}
\bar{S}^T\bar{D}rr^T\bar{D}\bar{S}=r^T\bar{D}r\bar{S}^T\bar{D}\bar{S}
\end{equation} which is
satisfied if and only if $r(\mathbf{e})=k\bar{S}$.
\end{proof}

\renewcommand{\proofname}{Proof of Theorem \ref{thm:stable}}
\begin{proof}
Consider the positive semidefinite function
\begin{equation}\label{eq:laypu_three}
V(\mathbf{e})=[r(\mathbf{e})-\tilde{s}^*(\mathbf{e})\bar{S}]^T[r(\mathbf{e})-\tilde{s}^*(\mathbf{e})\bar{S}]
\end{equation}
and its partial derivative with respect to $\mathbf{e}$
\begin{align}
  \frac{\partial V}{\partial \mathbf{e}}&=2\frac{\partial [r^T-\tilde{s}^*\bar{S}^T]}{\partial
  \mathbf{e}}[r-\tilde{s}^*\bar{S}]\nonumber\\
  &=2[\Lambda(\mathbf{e})-\frac{\partial \tilde{s}^*}{\partial
  \mathbf{e}}\bar{S}^T][r(\mathbf{e})-\tilde{s}(\mathbf{e})\bar{S}]\label{eq:dvde}
\end{align}
The partial derivative $\frac{\partial \tilde{s}^*}{\partial
\mathbf{e}}$ is calculated by first solving
\begin{equation*}
  \frac{\partial s_N}{\partial \mathbf{e}}= \Lambda(\mathbf{e}) (\bar{D}^T+\bar{D})
  r(\mathbf{e})
  \end{equation*}
  and \begin{equation*}
  \frac{\partial s_D}{\partial \mathbf{e}}=\Lambda(\mathbf{e}) \bar{D}^T \bar{S}
\end{equation*}
which then yield
\begin{equation}
\frac{\partial \tilde{s}^*}{\partial \mathbf{e}}=\frac{s_D
\Lambda(\mathbf{e}) (\bar{D}^T+\bar{D}) r(\mathbf{e})-s_N
\Lambda(\mathbf{e})\bar{D}^T \bar{S}}{s_D^2}\triangleq N(e)
\end{equation}

Hence the derivative of $V(\mathbf{e})$ with respect to $t$ is
\begin{align}
  \frac{d}{dt}V(\mathbf{e})&=(\frac{\partial V(\mathbf{e})}{\partial \mathbf{e}})^T \frac{d
  \mathbf{e}}{dt}\nonumber\\
  &=-2[r(\mathbf{e})-\tilde{s}^*(\mathbf{e})\bar{S}]^T[\Lambda(\mathbf{e})-N(\mathbf{e})\bar{S}^T]^T \hat{H} \hat{H} ^T \Lambda(\mathbf{e})
M(\mathbf{e})^T[r(\mathbf{e})-\tilde{s}^*(\mathbf{e})\bar{S}]\label{eq:mydvdt_temp}
\end{align}

 When
\begin{equation}\label{eq:myMM}
  M(\mathbf{e})=s_D^2 I_3-s_D \bar{S} r(\mathbf{e})^T(\bar{D}^T+\bar{D}) + s_N
  \bar{S} \bar{S}^T \bar{D}
\end{equation}
the derivative of $V(\mathbf{e})$ in \eqref{eq:mydvdt_temp} is
negatively semi-definite with the expression
\begin{equation}
\dot{V}(\mathbf{e})=-\frac{2}{s_D^2}[r(\mathbf{e})^T-\tilde{s}^*(\mathbf{e})\bar{S}^T]M(\mathbf{e})\Lambda(\mathbf{e})^T\hat{H}\hat{H}
^T\Lambda(\mathbf{e})M(\mathbf{e})^T[r(\mathbf{e})-\tilde{s}^*(\mathbf{e})\bar{S}]
\end{equation}
Indeed, $V(\mathbf{e})$ is a valid Lyapunov function candidate.

For the autonomous system when the derivative of the candidate
Lyapunov function is negative semi-definite, the asymptotic
stability is concluded based on the powerful invariant set theory.

For the negative semi-definite function $V(\mathbf{e})$, there is an
invariant set
\begin{equation*}
  \Omega_\sigma=\{\mathbf{e}\in \text{Im}\hat{H}:V(\mathbf{e})\leq c,c\in\mathbb{R}^+\}
\end{equation*}
The set of points in $\Omega_\sigma$ where $\dot{V}(\mathbf{e})=0$
satisfies both conditions
\begin{equation}\label{eq:invariantsetCondition_1}
\bar{S}^T\bar{D}r(\mathbf{e})\neq 0
\end{equation}
and
\begin{equation}\label{eq:temp3}
  \hat{H}^T\Lambda(\mathbf{e})M(\mathbf{e})^T[r(\mathbf{e})-\tilde{s}^*(\mathbf{e})\bar{S}]=0
\end{equation}
where condition \eqref{eq:invariantsetCondition_1} is always true
under Assumption \ref{assumption_generic}.

When the three agents are connected over the graph shown in Fig.
\ref{fig:three_topo}, matrix $\hat{H}^T\Lambda(\mathbf{e})\in
\mathbb{R}^{6\times 3}$ has rank $3$ for all
$\mathbf{\mathbf{e}}=\hat{H}\mathbf{z}\in\mathbb{E}$.

In order to find the largest invariant set, the singularity of
matrix $M(\mathbf{e})$ is crucial to the asymptotic stability. As
pointed out by Lemma \ref{lemma:M_nonsigular}, $M(\mathbf{e})$ is a
full rank matrix when $r(\mathbf{e})\neq k\bar{S}$. Thus equation
\eqref{eq:temp3} is satisfied if and only if there exist
$\mathbf{e}_f$ such that
$r(\mathbf{e}_f)=\tilde{s}^*(\mathbf{e}_f)\bar{S}$. Thus
\begin{equation}\label{eq:invariantSet_2}
  \mathcal{I}_{\mathbf{e}}=\{\mathbf{e}\in\text{Im}\hat{H}:r(\mathbf{e})-\tilde{s}^*_f\bar{S}=0\}
\end{equation}
where $\tilde{s}^*_f=\tilde{s}^*(\mathbf{e}_f)$ is the largest
invariant set for dynamic system \eqref{eq:mydes_M}.

If we let $\epsilon$ being the smallest eigenvalue of $RMM^TR^T$
during the entire convergence, the derivative of $V(\mathbf{e})$ is
bounded by $$\dot{V}(\mathbf{e})\leq -
\frac{2\epsilon}{\tilde{s}_D^2}
\|r(\mathbf{e})-\tilde{s}(\mathbf{e})\bar{S}\|^2: \triangleq -\theta
$$
which further yields $V(\mathbf{e})\leq V(\mathbf{e}_0)
e^{-\theta}$. According to some trivial calculations, we conclude
that $\|r(\mathbf{e})-\tilde{s}(\be) \bar{S}\|$ exponentially
converges to zero.
\end{proof}
\renewcommand{\proofname}{PROOF}

According to $\mathcal{I}_{e}$, once the formation system forms a
geometry in $\{\be|S\}$, it stays there from then on.

Equation \eqref{eq:mydynS} is a nonlinear map $\tilde{s}\in
C(\mathbb{E}): \mathbb{E} \rightarrow \mathbb{R}$ where given an
assigned shape $S$, the
 initial geometry  $\mathbf{e}_0$ in the nonlinear control law
\eqref{eq:mydynS} uniquely determines the stable scale
$\tilde{s}^*_f$.  However, due to the nonlinearity,  it is difficult
to predict the final stable value $\tilde{s}^*_f$.

On the other hand, although the exact value of $\tilde{s}^*_f$ is
what we are seeking for, we would like to discuss the
controllability of the algorithm when the scale is expected to
converge to some arbitrary fixed value. In such a case, the problem
is recast into a general formation control problem with a specified
desired geometry, which is explored in quite a few works, e.g.,
\cite{Belta04optimal}. However, we believe it is still necessary to
make this extension so as to broaden the applications of the
algorithm. This problem is covered in Subsection \ref{sec:multiple}
after we extend the nonlinear control law to multiple($n>3$) agents
case.

\subsection{Formation with multiple agents}
Some of related literatures concerning formation control were
restricted to three agents\cite{Basiri10angleonly,Bhatt09geometric}
and left the multiagents case as the future work. Extending
algorithms into multiple agents case would have to deal with the
complexity issue where the primary one is the selection of the
underlying graph. A graph with three nodes connected with each other
is a quite special case: it is a complete graph, a minimally rigid
graph and a ring graph. During the exposition of Theorem
\ref{thm:stable}, based on the cosine law, this particular property
of the three nodes ensures the compact form of the partial
derivative of the value function $L$. However, when the number of
agents exceeds three, the cosine law dose not always applies.

In order to obtain the nonlinear control law as in three agents
formations, an intuitive idea is to  adopt triangle as the basic
unit of the underlying graph. We define the \emph{triangular
complement} of a graph which would lead to some interesting and
convenient results.
\begin{defn}
  The \emph{triangular complement} of a graph $G=(V,E)$ is a graph $G'=(V,E')$
  with the same
  vertex set as $G$ and node $i$ are connected to node $j$ in $G'$ if and only if
  the minimal distance between $i,j$ in $G$ are two.
   The graph sum $G_\triangle=G+G'$
  has vertex set $V$ and edges $E_\triangle=E\cup E'$.
\end{defn}
\begin{rm}
  When the triangular complement graph is added to the minimally rigid
graph, its edges are labeled from $|E|+1$ to $|E|+|E'|$.
\end{rm}
\begin{cor}
  The graph sum $G_\triangle=G+G'$ equals to
  $G^2$.
\end{cor}
For the definition of $G^2$, please refer to, for example,
\cite{Diestel10,Anderson09easilyLocalizable}.

\begin{figure}
 \subfigure[$G_1$]{\label{fig:minimalandComplement_a}
\begin{minipage}[b]{0.23\linewidth}
\centering
\includegraphics[scale=0.5]{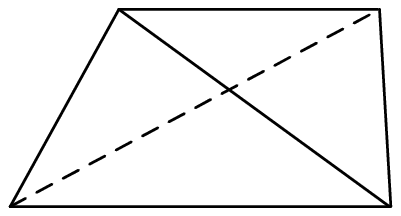}
\end{minipage}}
\subfigure[$G_2$]{\label{fig:minimalandComplement_b}
\begin{minipage}[b]{0.23\linewidth}
\centering
\includegraphics[scale=0.5]{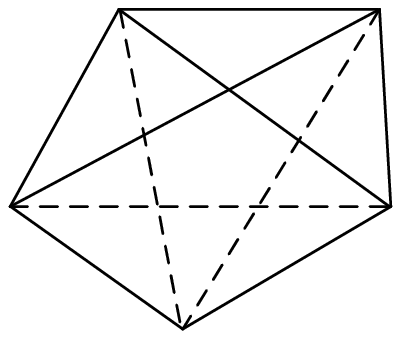}
\end{minipage}}
\subfigure[$G_3$]{\label{fig:minimalandComplement_c}
\begin{minipage}[b]{0.23\linewidth}
\centering
\includegraphics[scale=0.5]{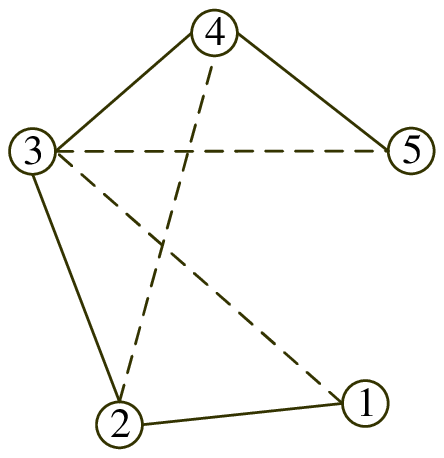}
\end{minipage}}
\subfigure[$G_4$]{\label{fig:minimalandComplement_d}
\begin{minipage}[b]{0.23\linewidth}
\centering
\includegraphics[scale=0.5]{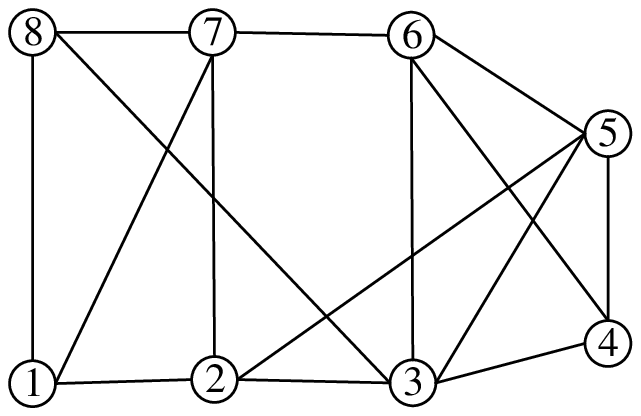}
\end{minipage}}
\caption{Graphs and their triangular complement graphs shown with
dashed lines. $G_1$ and $G_2$ are minimally rigid graphs, $G_3$ is
only a connected graph and $G_4$ is a globally rigid
graph.}\label{fig:example_triangular}
\end{figure}

 The
triangular complement graph $G'$ contains the third edges that
constitute triangles with the neighboring edges in $G$, as shown in
 Fig. \ref{fig:minimalandComplement_a} and Fig.
\ref{fig:minimalandComplement_b} where the solid lines are edges in
$G$ and the dashed lines belong to $G'$. However, the neighboring
edges from $G$ and $G'$ respectively do not
  necessarily have a third edge in $G_\triangle$ that form a
  triangle, as shown in Fig. \ref{fig:minimalandComplement_c}.

It is not necessarily true that $G_\triangle$, or equivalently
$G^2$, being a complete graph even when $G$ is minimally rigid or
globally rigid, as for instance the situation in Fig.
\ref{fig:minimalandComplement_d} where the length between node 4 and
node 1 is greater than 2. That is the reason we name it triangular
complement graph rather than the general complement graph as in
\cite{Diestel10}.

The three edges $e_i$, $e_j$ and $e_\gamma$ in $G_\triangle$ that
form a triangle is denoted by $\triangle_{ij\gamma}$. We have made
the assumption in the preliminary  that each agent has the knowledge
of the coordinates of its neighbors, thus the agent that is adjacent
to $e_i$ and $e_j$ in $G$ is able to calculate the coordinates of
the edges $e_\gamma$ in $G'$, i.e., $e'_\gamma=e_i+e_j$ or
$e'_\gamma=e_i-e_j$ for $\triangle_{ij\gamma}$.

For multiple agents case, the parallel form of matrix $\bar{D}$ in
\eqref{eq:D_three} is denoted by $\hat{D}(G_\triangle)\in
\mathbb{R}^{|E_\triangle|\times |E_\triangle|}$  with entries
\begin{equation*}
 \hat{d}_{ij}=\begin{cases}
\sum\limits_{k:k\in\mathcal{N}_e(i)} \bar{s}_k-4\bar{s}_i& ,i=j\\
\bar{s}_j-\bar{s}_{\gamma} &, \triangle_{ij\gamma},i,j\in E,\gamma \in E\\
\bar{s}_j &, \triangle_{ij\gamma},i,j\in E, \gamma \in E'\\
-\bar{s}_\gamma &, \triangle_{ij\gamma}, i,\gamma\in E, j\in E'\\
 0 &, \text{others}
\end{cases}
\end{equation*}
If we define an expended shape vector $\hat{S}=[\bar{S}
;\mathbf{0}]\in \mathbb{R}^{|E_\triangle|}$ and similarly
$\hat{r}(\mathbf{e})=[r(\mathbf{e});r'(\mathbf{e})]$,
$\hat{\mathbf{e}}=[\mathbf{e};\mathbf{e}']$ where $r'(\mathbf{e})$,
the norm of edges in $G'$, is calculated by the corresponding agent
using local coordinates of the other two edges in $G$, and
consequently, similar to \eqref{eq:mydynS} the trajectory of the
scale function in terms of $\be$ is
\begin{equation}\label{eq:dynamicS_multiple}
  \hat{s}^*(\mathbf{e})=\frac{\hat{r}(\mathbf{e})^T\hat{D}\hat{r}(\mathbf{e})}{\hat{S}^T\hat{D}\hat{r}(\mathbf{e})}:\triangleq
  \frac{\hat{s}_N}{\hat{s}_D}
\end{equation}
and
\begin{equation*}
  \frac{\partial \hat{s}_N}{\partial
  \mathbf{e}}=\hat{\Lambda}(\mathbf{e})(\hat{D}+\hat{D}^T)\hat{r}(\mathbf{e})
\end{equation*}
\begin{equation*}
   \frac{\partial \hat{s}_N}{\partial
  \mathbf{e}}=\hat{\Lambda}(\mathbf{e})\hat{D}^T \hat{S}
\end{equation*}
where $\hat{\Lambda}(\mathbf{e})=\frac{\partial
\hat{r}(\mathbf{e})}{\partial \mathbf{e}}=\begin{bmatrix}
\Lambda(\mathbf{e})&\Lambda'(\mathbf{e})
\end{bmatrix}\in \mathbb{R}^{2|E|\times |E_\triangle|}$.
This further yields
\begin{equation}\label{eq:M_multiple}
  \hat{M}(\mathbf{e})=\hat{s}_D^2\begin{bmatrix}I_{2n-3}
  &0\end{bmatrix}-\hat{s}_D\hat{S}\hat{r}^T(\mathbf{e})(\hat{D}^T+\hat{D})+\hat{s}_N\hat{S}
  \hat{S}^T\hat{D}
\end{equation}
where $\hat{H}$ is still the oriented incidence matrix of $G$.

The derivative of the Lyapunov function \eqref{eq:laypu_three} is
then
\begin{equation}\label{eq:dvde_multi}
\dot{V}(\mathbf{e})=-\frac{2}{\hat{s}_d^2}[r(\mathbf{e})^T-\hat{s}(\mathbf{e})\bar{S}^T]
\hat{M}(\mathbf{e})\hat{\Lambda}(\mathbf{e})^T\hat{H} \hat{H}
^T\hat{\Lambda}(\mathbf{e})\hat{M}(\mathbf{e})^T[r(\mathbf{e})-\hat{s}(\mathbf{e})\bar{S}]
\end{equation}

\begin{thm}\label{thm:stable_multi}
The formation system
\begin{equation}\label{eq:myopt_multi} \dot{\mathbf{e}}=-\hat{H}\hat{H}^T\hat{\Lambda}(\mathbf{e})
\hat{M}(\mathbf{e})^T[r(\mathbf{e})-\hat{s}^*(\mathbf{e})\bar{S}],
\mathbf{e}_0\in \Omega_\sigma
\end{equation}
over a minimally rigid graph is exponentially stable and converges
to the largest invariant set $\mathcal{I}_e$ in
\eqref{eq:invariantSet_2} if $\hat{s}^*(\mathbf{e})$ and $\hat{M}$
are given in \eqref{eq:dynamicS_multiple} and \eqref{eq:M_multiple}
respectively.
\end{thm}
\begin{proof}
The first part of the stability proof is identical to that of three
agents situation. When $n\geq 4$, for a minimally rigid graph we
have the similar results that $\text{rank} \hat{M}(\mathbf{e})=2n-3$
if and only if $r(\mathbf{e})=k\bar{S}$, and it is also trivial that
$\hat{H}^T\Lambda(\mathbf{e})\hat{M}^T$ is of rank $2|V|-3$.
Following the proof of Theorem \ref{thm:stable} we conclude that the
largest invariant set is $\mathcal{I}_{\mathbf{e}}$ in
\eqref{eq:invariantSet_2}, which indicates that the formation system
converges to the desired shape at some scale $\hat{s}_f$ for a
sufficiently small $\sigma$.

The proof of exponentially convergence is consistent to that of
three agents case and is omitted here.
\end{proof}
\begin{rem}
When we considers only three agents in a formation system, the
underlying graph is a globally rigid one and a minimally rigid one
simultaneously, thus the requirements on the initial realization is
relaxed. However for system with up to four agents, this privilege
dose not holds true. Thus it is required that the initial geometry
$\be_0\in\Omega_\sigma$, and Remark \ref{rem:initial} applies.
\end{rem}
\begin{rem}
  If we look at $[r(\mathbf{e})-\hat{s}^*(\mathbf{e})\bar{S}]$ in the control law,
  we require only a local
  coordinate for each node rather than a global one.
  The reason is that in the control law, we only concern with the
  relative distance between two agents, as represented by vector
  $r(\mathbf{e})$. On the other hand, in order to calculate $r'(\mathbf{e})$,
  local coordinates of two neighboring edges would be enough as well.
  However, for the rigidity matrix $\hat{R}(\mathbf{e})=\hat{\Lambda}(\mathbf{e})^T\hat{H}$,
  it requires the relative coordinates in a common reference frame. So for convenience,
  we assign a global coordinate for the formation system.
\end{rem}

With the time-varying scale function $\hat{s}^*$, the desired
geometry $\hat{s}(\be_{f})$ is adjusted according to the current
geometry. Equation \eqref{eq:dynamicS_multiple} is a
nonlinear mapping $\hat{s}\in C(\mathbb{E}): \mathbb{E} \rightarrow
\mathbb{R}$ where the
 initial condition  $\mathbf{e}_0$ uniquely determines the stable scale
$\hat{s}^*_f$ for the prescribed shape $S$. The scale function
$\hat{s}^*$ would lead to a bounded scale and ensure a relatively
small cost value. This performance is observed in the experiments in
Section \ref{sec:exp}.

\subsection{Controllability of the multi-agent formation
system}\label{sec:multiple} The aforementioned nonlinear functions
$\tilde{s}^*$ and $\hat{s}^*$ put us in a blind position about the
exact side lengths when agents get stabilized, and thus restricts
the applications of the algorithms. For instance, in the task of
payload transport, the desired geometry is constrained within a
certain area determined by the  cargo. Formation that shrinks to
some geometry with a relatively small scale  around the center of
gravity may result in a sensitive balance system and a sparse
formation where agents locate at the edge of the cargo may easily
failed.

In order to drive the formation system to converge to an assigned
geometry that belongs to $\{\mathbf{e}|S\}$, we add an additional
control gain $\Pi$ to the nonlinear control law
\begin{equation}\label{eq:mydzs_2}
  \dot{\bar{\mathbf{z}}}=-\Pi\hat{H}^T \hat{\Lambda}(\bar{\mathbf{e}})
  \hat{M}(\bar{\mathbf{e}})^T [r(\bar{\mathbf{e}})-\hat{s}^*(\bar{\mathbf{e}})\bar{S}]
\end{equation}
where $\bar{\mathbf{e}}\in\mathbb{E}$, $\bar{\mathbf{z}}\in
\mathbb{Z}$ and $\Pi=\text{diag}(a_1,a_2,\ldots,a_{2n})$. The edge
and the state of each agent that evolve along \eqref{eq:mydzs_2} are
denoted by $\bar{\mathbf{e}}$ and $\bar{\mathbf{z}}$ respectively.

The weight factor $a_i$ adjusts the convergence rate of each edge
and thus tunes the stable value $\bar{\mathbf{e}}_f$ and
$\hat{s}^*(\bar{\mathbf{e}}_f)$, and we will prove that  the
formation system under control law \eqref{eq:mydzs_2} is
controllable.
\begin{thm}
  Given a desired geometry $\bar{\mathbf{e}}_f$ in $\{\mathbf{e}|S\}$, there exists $\Pi$ satisfying
 $a_i>0, \forall i\in [1,2n]$ in \eqref{eq:mydzs_2} such that a formation
  system over a minimally rigid underlying graph exponentially converges to $\bar{\mathbf{e}}_f$ given that
  $\bar{\mathbf{e}}_0\in\Omega_\sigma$.
  \end{thm}

\begin{proof}
    We prove the theorem by finding the appropriate $\Pi$ with respect to $\lambda$
  such that $r(\bar{\mathbf{e}})\rightarrow \lambda \bar{S}$.
The procedure could be carried out by taking the original system
\eqref{eq:mydes_M} as a reference system.

Assume $r(\mathbf{e})\rightarrow \hat{s}^*(\mathbf{e}_f)\bar{S}$ and
$r(\bar{\mathbf{e}})\rightarrow
\hat{s}^*(\bar{\mathbf{e}}_f)\bar{S}$ where $
\hat{s}^*(\bar{\mathbf{e}}_f)=\lambda$, and
$\mathbf{e}_0=\bar{\mathbf{e}}_0$. Let
$k=\hat{s}^*(\bar{\mathbf{e}}_f)/\hat{s}^*(\mathbf{e}_f)$.

  Apparently, $\dot{\bar{\mathbf{e}}}=\hat{H} \Pi \dot{\mathbf{z}}$. Integrating on each
  side of the equation yields
  \begin{equation}\label{eq:mytemp1}
    \bar{\mathbf{e}}_f-\bar{\mathbf{e}}_0=\hat{H} \Pi (\mathbf{z}_f-\mathbf{z}_0)
  \end{equation}
  The condition that $r(\bar{\mathbf{e}})\rightarrow k \hat{s}^*\bar{S}$ is equivalent
  to, for each edge vector $e_i$,
  \begin{equation}\label{eq:epieandz}
    \bar{\mathbf{e}}_{f_i}=kR(\theta){\be_f}_i+p_i
  \end{equation}
  with vector $\mathbf{p}=[p_i]\in \mathbb{E}$.
  Substituting \eqref{eq:epieandz} into \eqref{eq:mytemp1} we
  obtain
  \begin{equation}\label{eq:myfindL-1}
   \hat{H} \Pi
   (\mathbf{z}_f-\mathbf{z}_0)=k\hat{R}(\theta)\hat{H} \mathbf{z}_f-\mathbf{e}_0+\mathbf{p}
  \end{equation}
  where $\hat{R}(\theta)$ is a block diagonal matrix with $2\times
  2$ identical diagonals $R(\theta)$.
  In order to analysis the solution of $a_i$, \eqref{eq:myfindL-1}
  could be rewritten into a standard linear matrix equation  with respect to
  $a_i$
\begin{align}\label{eq:myfindPi-final}
  \hat{H} \Lambda(\Delta \mathbf{z})\begin{bmatrix}
    a_1\\a_2\\ \vdots \\a_{2n}
  \end{bmatrix}&=k\hat{R}(\theta)\hat{H}\mathbf{z}_f-\mathbf{e}_0+\mathbf{p}\nonumber\\
&\triangleq kO(\mathbf{z}_f)-\mathbf{e}_0+\mathbf{p}
\end{align}
where $\Lambda(\Delta \mathbf{z})=\text{diag}
({\mathbf{z}_f}_i-{\mathbf{z}_0}_i)$.
%

Noting that for a minimally rigid graph, $\text{rank}(\hat{H}
)=2n-2$, i.e. $\text{dim}(\text{ker}(\hat{H}))=2$. On the other
hand, $\text{rank}(O)=2(2n-3)$. We can always find a set of positive
numbers $a_i$ in \eqref{eq:myfindPi-final} by choosing appropriate
$\mathbf{p}$ such that the number of the nonzero entries of the
right hand side vector is less than $2n-2$.

%
%

With the positive control gain $\Pi$, the stability of the formation
system under the advanced control law \eqref{eq:mydzs_2} could be
proved along the same line as Theorem \ref{thm:stable_multi}, which
finishes the proof.
\end{proof}
\begin{rem}
  In order to ensure the convergence to the desired shape,
  restrictions on the initial conditions, i.e., Remark
  \ref{rem:initial} remain applies.
\end{rem}

Recall Theorem \ref{thm:JvJc}, under the endpoint constraints
  $\hat{s}^*_f=\lambda$, the control law $\eqref{eq:mydzs_2}$ still inherits
the suboptimality of $J_v|_{\Theta=\hat{s}^*}$ with respect to
$J_c$.

\begin{figure}
\begin{center}
\includegraphics[scale=0.8]{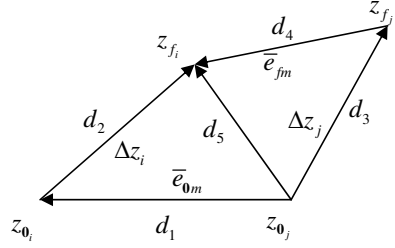}
\caption{Geometric analysis of the control gain $\Pi$}
\label{fig:controllability}
\end{center}
\end{figure}

The existence of the positive gain $\Pi$ could also be validated by
the geometric representation of \eqref{eq:myfindPi-final}. Condition
\eqref{eq:myfindPi-final} is satisfied if and only if for all
$m\in|E_\triangle|$,
\begin{equation}\label{eq:controllability_geometry}
  \begin{bmatrix}
    a_{2i-1}&0\\0&a_{2i}
  \end{bmatrix}\Delta z_i
  -
\begin{bmatrix}
    a_{2j-1}&0\\0&a_{2j}
  \end{bmatrix}\Delta z_j
  =
  \bar{e}_{f_m}-\bar{e}_{0_m}
\end{equation}
where $\Delta z_i={z_f}_i-{z_0}_i$. Equation
\eqref{eq:controllability_geometry} determines the adjacency of
vectors in Fig. \ref{fig:controllability}. According to the triangle
inequality,
$$d_2-d_1-d_3\leq d_4 \leq d_1+d_2+d_3$$ where
$d_1=\|{e_0}_1\|$, $d_2=\|l_i({z_f}_i-{z_0}_i)\|$,
$d_3=\|l_j({z_f}_j-{z_0}_j)\|$, $d_4=\lambda s_m$ and
$l_i=\text{diag}(a_{2i-1},a_{2i})$. By positioning vector
$\bar{e}_{f_m}$ on the upper right side of $\bar{e}_{0_m}$, it is
always guaranteed that $l_i>0,\forall i\in[1,n]$.

The discussion on the controllability allows the algorithms to be
applied to situation that requires the switching of the geometry's
scale during the mission. For example when underwater vehicles are
mapping out oceanbed\cite{Kalantar08scale}, the scale of the
formation is adjusted online to decrease the interpolation error.

As stated before, this special problem we considered above falls
into to the general
 formation control category where the optimization is carried out under the constraint of a fixed
 geometry. However those existing results mainly focus purely on
 minimal energy expenditure\cite{Belta04optimal}
  rather than the geometrical performance during the process. In our research the cost
  function is carefully designed so as to have the group of agents
  exhibits prominent geometrical performance during convergence.

  In reality, the final scale $\hat{s}^*_f$ should be restrained within an interval. Generally, the
  upper bound could be the communication
  range of agents, and for lower bound, the collisions of agents should be taken into
  account.

In this section, motivated by the fact that a scale being adjusted
online may result in better cooperative performance, we found a
time-varying scale function in terms of the edge vector $\mathbf{e}$
that further reduces the global minimum of system with a constant
scale. By introducing the triangular complement graph, we derived a
compact form of the nonlinear control law for multiple agents. With
the additional control gain, the nonlinear control law is also
applicative to general formation control problems with fixed final
geometry.

\section{Geometrical performance in sensor-target localizations}\label{sec:localization}
 The localization of target in sensor networks is always carried out by a fleet of
 three
UAVs in triangle. Each of the UAVs is equipped with sensors that
gather information of the target based on bearing
measurements\cite{Dovganccay05bearing}, range
measurements\cite{Martinez06rangelocalization} or scan
measurements\cite{Dogancay07scan}. A lot of literatures are devoted
to the exploit of the sensors positioning strategy as it is strongly
related to the localization performance. Based on Fisher information
matrix, the pioneering work demonstrated in \cite{Bishop10sensor}
proved that sensors in a equilateral triangle provide the minimal
variance estimation of the target.

\begin{figure}
 \subfigure[Equilateral]{\label{fig:optimal_triangles_equa}
\begin{minipage}[b]{0.31\linewidth}
\centering
\includegraphics[scale=0.3]{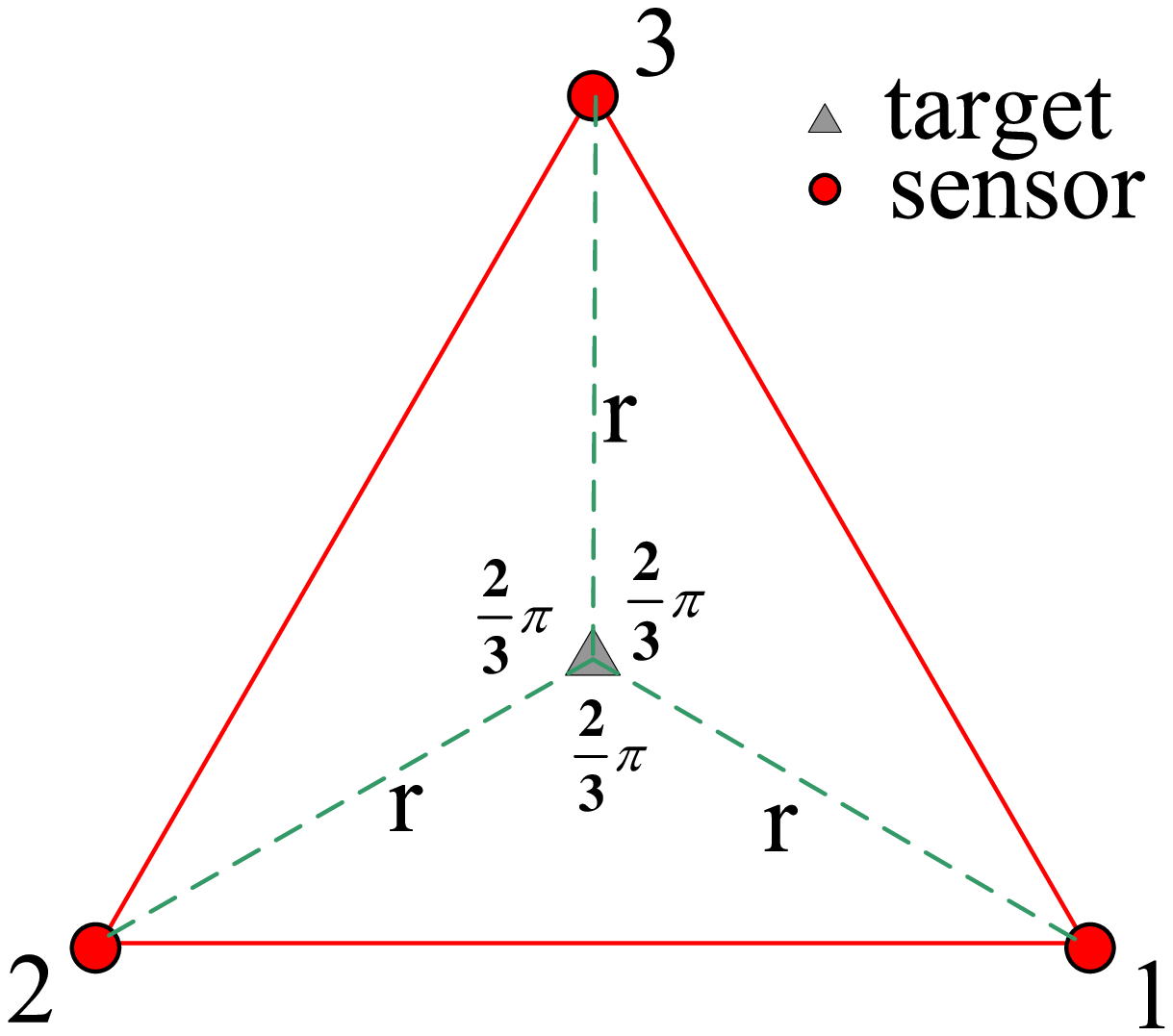}
\end{minipage}}
\subfigure[Isosceles]{\label{fig:optimal_triangles_isos}
\begin{minipage}[b]{0.31\linewidth}
\centering
\includegraphics[scale=0.3]{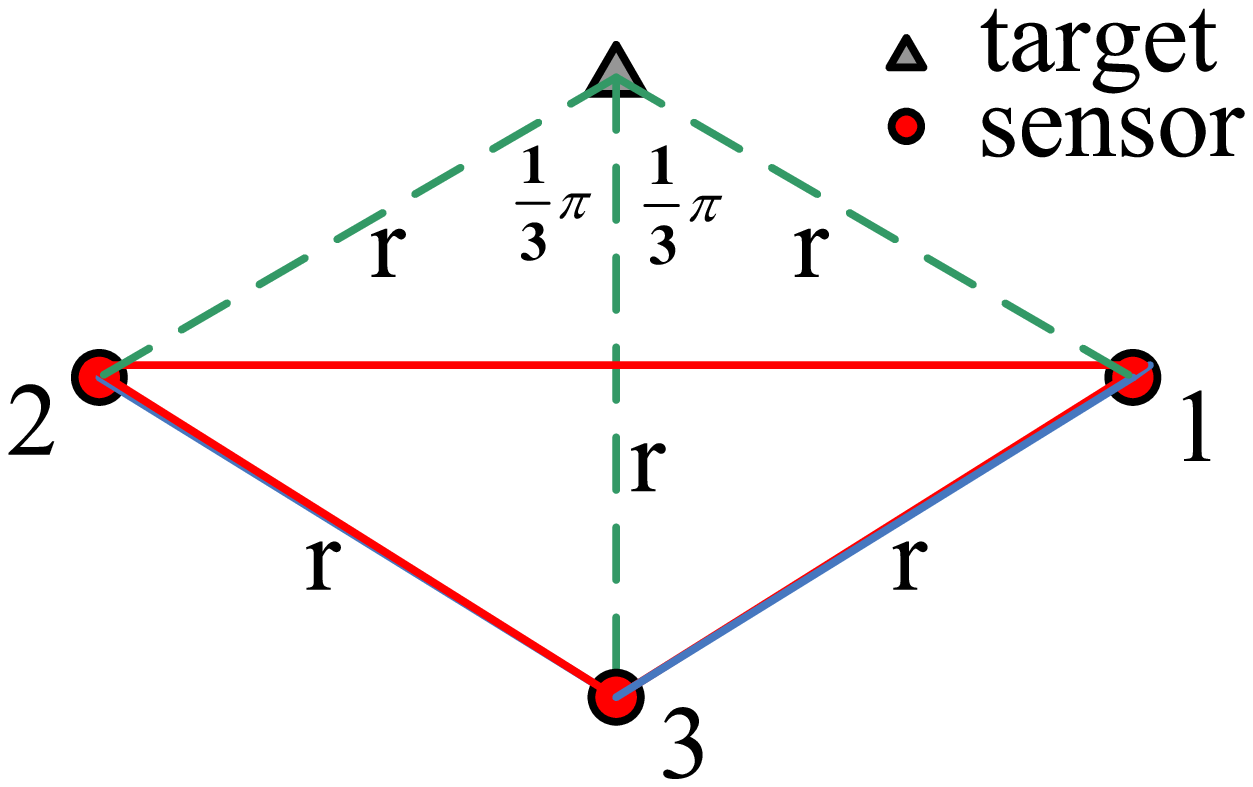}
\end{minipage}}
\subfigure[Scalene]{\label{fig:optimal_triangles_similar}
\begin{minipage}[b]{0.31\linewidth}
\centering
\includegraphics[scale=0.3]{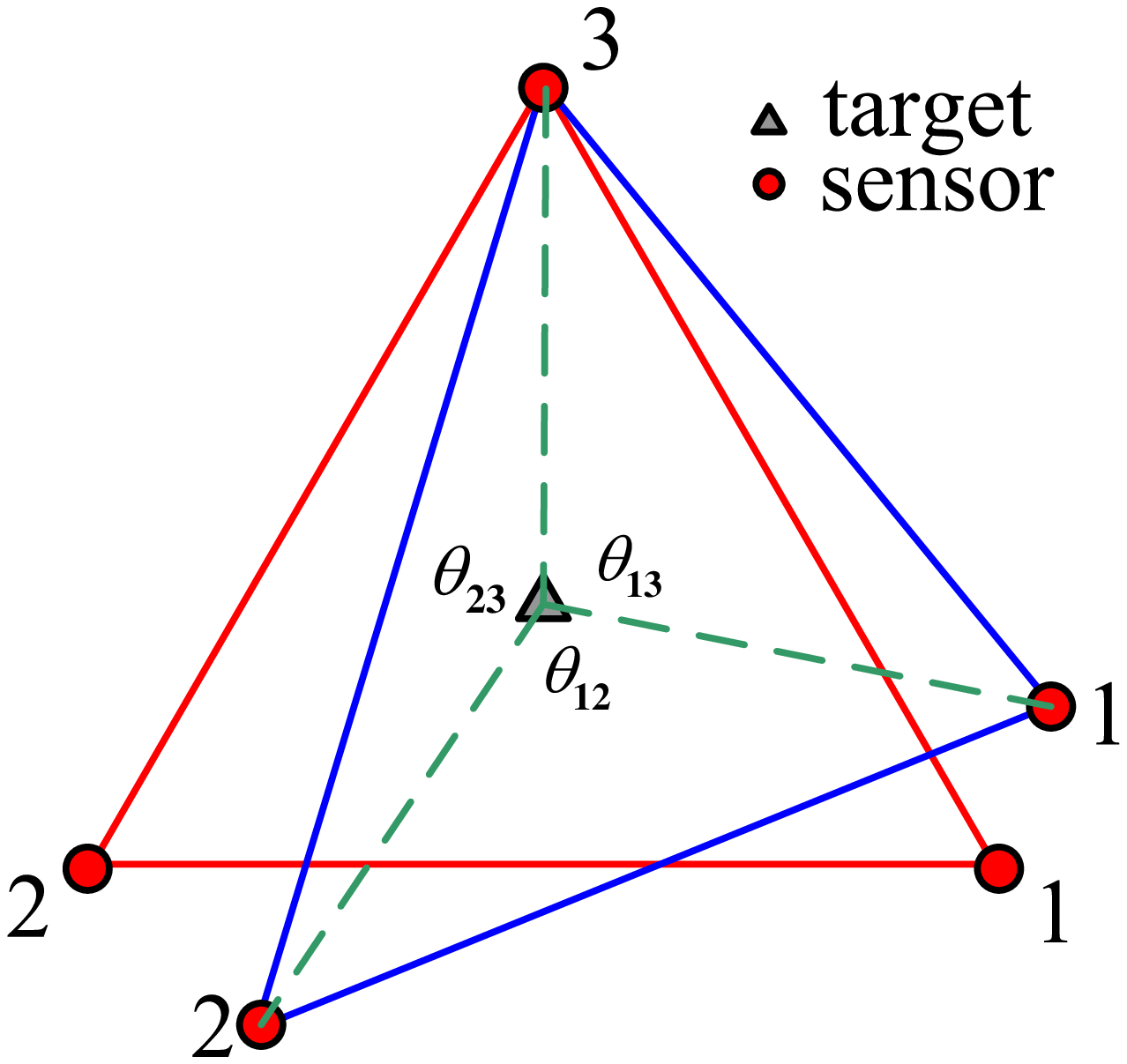}
\end{minipage}}
\caption{Optimal triangles for bearing-only sensor-target
localization and a normal triangle as shown in color
blue}\label{fig:optimal_triangles}
\end{figure}

Consider the special scenario where sensors (or UAVs that are
equipped with sensors) are deployed on a circle of radius $r$ around
the target. When $r$ is sufficiently large such that the distances
from the target to the three sensors could be considered as being
constant during the entire formation attainment process, one of the
main results in \cite{Bishop10sensor} could be restated specifically
as
\begin{lem}
  Consider angle-based bearing-only localization with three sensors. Assume the distance between sensor $i$ and the
  target, denoted by $r_i$, satisfies $r_i\equiv r,\forall i=1,2,3$.
  Then the variance estimation of the target position is minimized
  if the triangle formed by the three sensors satisfies
  \begin{align}
    &\theta_{12}=\theta_{13}=\frac{1}{2}\arccos(-\frac{1}{2})\nonumber\\
    &\theta_{23}=2\pi
    -\theta_{12}-\theta_{13}\label{eq:localization_shape}
  \end{align}
  where  $\theta_{ij}\in[0,\pi]$ is the angle subtended at the target by two sensors $i$ and
  $j$. Moreover, reflecting a sensor about the target position does not
affects the optimality of the geometry.
\end{lem}
When  the distances between sensors and the target are congruent,
the optimal geometries generated from \eqref{eq:localization_shape}
have only angles constrains and are divided into two subsets where
geometries in each of them are similar ones. Examples of the optimal
geometries in the two subsets are shown in Fig.
\ref{fig:optimal_triangles_equa} and Fig.
\ref{fig:optimal_triangles_isos} respectively. For situations in
Fig. \ref{fig:optimal_triangles_equa}, any equilateral triangles
with the target locate at the geometry center are considered to be
the optimal deployment of sensors while irrespective of their exact
sizes. Thus here we focus on this optimal case to exploit the
relationship between the variance estimation of the target and the
resemblance of a geometry with respect to the equilateral triangle.

Consider a target that lies close to an equilateral triangle whose
angles are denoted by $\theta'_{12}$, $\theta'_{13}$ and
$\theta'_{23}$ respectively, as shown in Fig.
\ref{fig:optimal_triangles_similar}. Without lose of generality,
assume $\theta'_{23}>\theta'_{12}>\theta'_{13}$. Then for simplicity, the
difference between a shape with respect to the
equilateral triangle could be measured by
\begin{equation}\label{eq:similarity}
  \delta^{-1}=\Delta_1+\Delta_2:\triangleq
  |\theta'_{12}-\theta'_{13}|+|\theta'_{23}-\theta'_{12}|
\end{equation}
A geometry with a larger $\delta$ has a higher degree of similarity
to a equilateral triangle.

The variance estimation of the  target position measured by the
sensors are evaluated by the determinant of the Fisher information
matrix $I(\theta)$. Under constraints \eqref{eq:localization_shape},
the determinant of $I(\theta)$ achieves its maximum
of\cite{Bishop10sensor}
\begin{equation}
  \det(I(\theta^*))=\frac{1}{\sigma_\theta^4}\sum\frac{3\sin^2
  \frac{2}{3}\pi}{r^4}=\frac{9}{4r^4\sigma^4}
\end{equation}
where $\sigma$ is the congruent error variance of sensors.

For a scalene triangle with angles $\theta'_{ij}$, the difference
between the determinant values of $\det(I(\theta^*))$ and
$\det(I(\theta'))$ is
\begin{equation}
  \Delta=\frac{1}{r^4\sigma^4}(\cos \theta'_{23}\cos(\theta'_{12}-\theta'_{13})+
  \cos \theta'_{13}\cos(\theta'_{12}-\theta'_{23})+ \cos \theta'_{12}(\theta'_{23}-\theta'_{13}))
\end{equation}

The following theorem shows how $\delta$ in \eqref{eq:similarity}
affects the localization accuracy.
\begin{thm}\label{thm:estimation_opt}
The following two conditions are true if
$\theta'_{13}>\frac{\pi}{2}$.
\begin{itemize}
\item[i).] when $\Delta_2$ is constant, $\Delta$
increases as $\Delta_1$ increases;
\item[ii).] when $\Delta_1$ is constant, $\Delta$ increases as
$\Delta_2$ increases.
\end{itemize}

\end{thm}
\begin{proof}
\begin{equation}
\frac{\partial \Delta}{\partial \Delta_1}=\frac{2}{3}(2\sin
\frac{2}{3}(2\pi-\Delta_2-2\Delta_1)-\sin\frac{2}{3}(2\pi-\Delta_2+\Delta_1))-\sin\frac{2}{3}(2\pi+2\Delta_2+\Delta_1))
\end{equation}
Similarly,
\begin{equation}
\frac{\partial \Delta}{\partial \Delta_2}=\frac{2}{3}(\sin
\frac{2}{3}(2\pi-\Delta_2+\Delta_1)+\sin\frac{2}{3}(2\pi-\Delta_2-2\Delta_1))-2\sin\frac{2}{3}(2\pi+2\Delta_2+\Delta_1))
\end{equation}
The sufficient conditions  for $\frac{\partial \Delta}{\partial
\Delta_1}>0$ and $\frac{\partial \Delta}{\partial \Delta_2}>0$ are
\begin{equation}\label{eq:temp4}\Delta_2+2\Delta_1< \frac{\pi}{2}\end{equation} when
$\Delta_1>\Delta_2$,
\begin{equation}\label{eq:temp5}0<\Delta_2-\Delta_1<\frac{\pi}{2},
2\Delta_1+\Delta_2<\frac{\pi}{2}\end{equation} when
$\Delta_1<\Delta_2$, and $\Delta<\pi/6$ when
$\Delta_1=\Delta_2=\Delta$. Thus irrespective of the relationship
between $\Delta_1$ and $\Delta_2$, $\frac{\partial \Delta}{\partial
\Delta_1}>0$ and $\frac{\partial \Delta}{\partial \Delta_2}>0$ are
sufficiently true if
$$0<\Delta_2-\Delta_1<\frac{\pi}{2}$$
and
$$0<2\Delta_1+\Delta_2<\frac{\pi}{2}.$$
or equivalently,
\begin{align}
  &\theta'_{23}+\theta'_{13}-2\theta'_{13}<\frac{\pi}{2}\nonumber\\
  &2\theta'_{23}-\theta'_{12}-\theta'_{13}<\frac{\pi}{2}
\end{align}
Thus under the assumption that
$\theta'_{23}>\theta'_{12}>\theta'_{13}$, inequality
$$\theta'_{13}>\frac{\pi}{2}$$
is the sufficient condition to $i)$ and $ii)$.
\end{proof}
It can be proved with some trivial calculation that the DOS
defined in \eqref{eq:distance} is positively monotonic to $\delta^{-1}$,
that is
\begin{cor}\label{cor:similarity}
  In bearing-only sensor-target localization with three sensors, assume the smallest angle that is
  subtended by two sensors
  at the target is greater than $\frac{\pi}{2}$.  Then the three
  sensors provides better estimation of the target location if the DOS of the triangle to the
   equilateral triangle is higher.
\end{cor}

Generally in UAV localizations, the three UAVs are set off from
certain locations and are commanded to fly towards the destination
separately. Localization of the target is carried out after they
attain the desired geometry at the destination. However, instead of
relying completely on the final attainment of the formation,
according to Theorem \ref{thm:estimation_opt}, we suggest that if we
can take the formation performance $J(\mathbf{e}_0,u,\Theta)$ into
account \emph{during} attainment, the geometries that are resemble to
the optimal one can provide pre-measurements of the target location
before arriving at the destination. Strategies could be made based
on those forecasted data and the executing time could be cut down if
the localization is not expected to be highly accurate.

Simulations are given at the end of Section \ref{sec:exp} to show
the improvement of the localization capability when the control law
\eqref{eq:u_varying} with $\tilde{s}(\be)=\tilde{s}^*(\be)$ is
applied.

\section{Examples and simulations}\label{sec:exp}
In order to validate the effectiveness and the system performance of
the proposed nonlinear control laws and the superiority of the
time-varying scale function, we compare system performance under the
control laws introduced in Section \ref{sec:inva} and Section
\ref{sec:varying}.

For multiple agents, the underlying  minimally rigid graph is not
unique, which inspires us to explore how the topology affects the
system performance. Recall that a triangular complement graph is
defined when applying the control law to multiple agents, we suggest
to categorize the minimally rigid graphs by the number of edges in
their
 triangular complement graphs, and then analysis their corresponding performance.

Apart from comparing the cost values, we also observe the maximal
and total distance agents travel. These might be important in
certain scenarios although they are not our primary concerns in this
research. However, this observation could be a starting point for
the future work.

The cooperative performance and its relationship to the cost
function is illustrated in an intuitive way  to further demonstrate
the feature of the cost function.

Conclusions and conjectures are drawn from the results on how the
underlying topology and the initial realizations are related to the
cooperative performance.

\subsection{The scales of geometries and the performance}
\begin{figure}
 \subfigure[Graph $G_a$]{\label{fig:four_topo_a}
\begin{minipage}[b]{0.45\linewidth}
\centering
\includegraphics[scale=0.4]{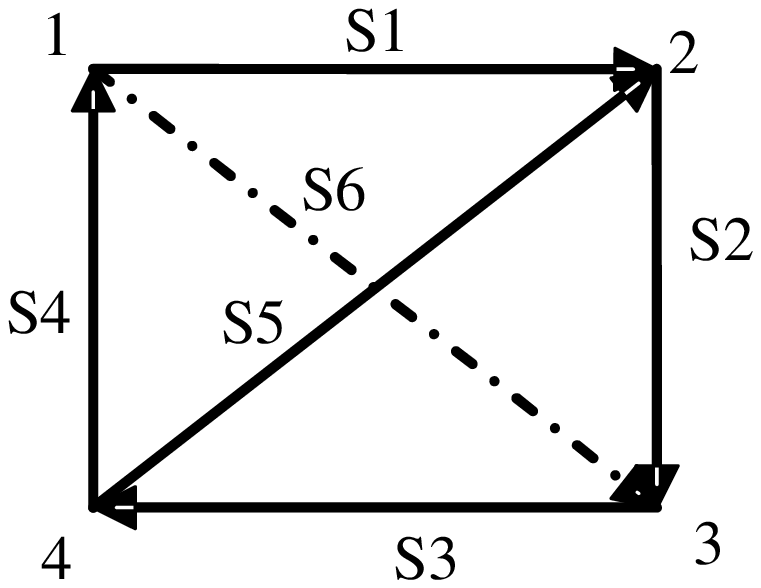}
\end{minipage}}
\subfigure[Graph $G_b$]{\label{fig:four_topo_b}
\begin{minipage}[b]{0.45\linewidth}
\centering
\includegraphics[scale=0.4]{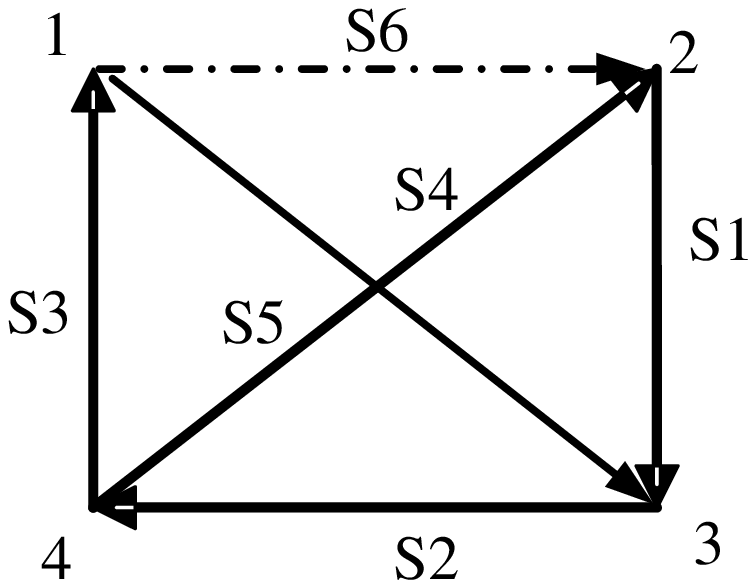}
\end{minipage}}
\caption{Two different graphs with the same number of triangular
complement edges}\label{fig:four_topo}
\end{figure}

\begin{figure}
 \subfigure[The cost value $J_c$]{\label{fig:time_invariant_s}
\begin{minipage}[b]{0.35\linewidth}
\centering
\includegraphics[width=2.8cm,height=2cm]{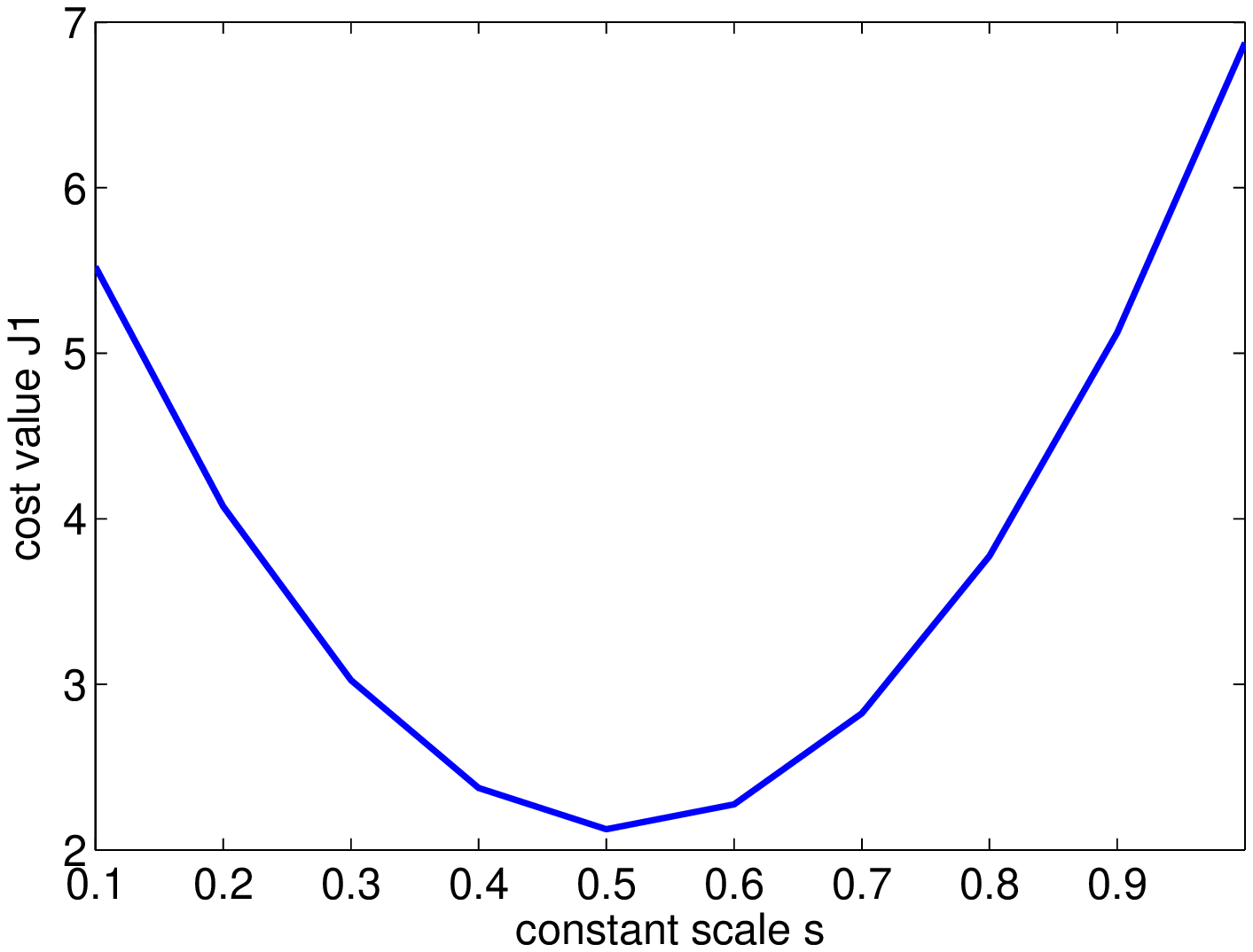}
\end{minipage}}
\subfigure[$s_c=0.5$]{\label{fig:invariant_s_05}
\begin{minipage}[b]{0.28\linewidth}
\centering
\includegraphics[width=2.8cm,height=2cm]{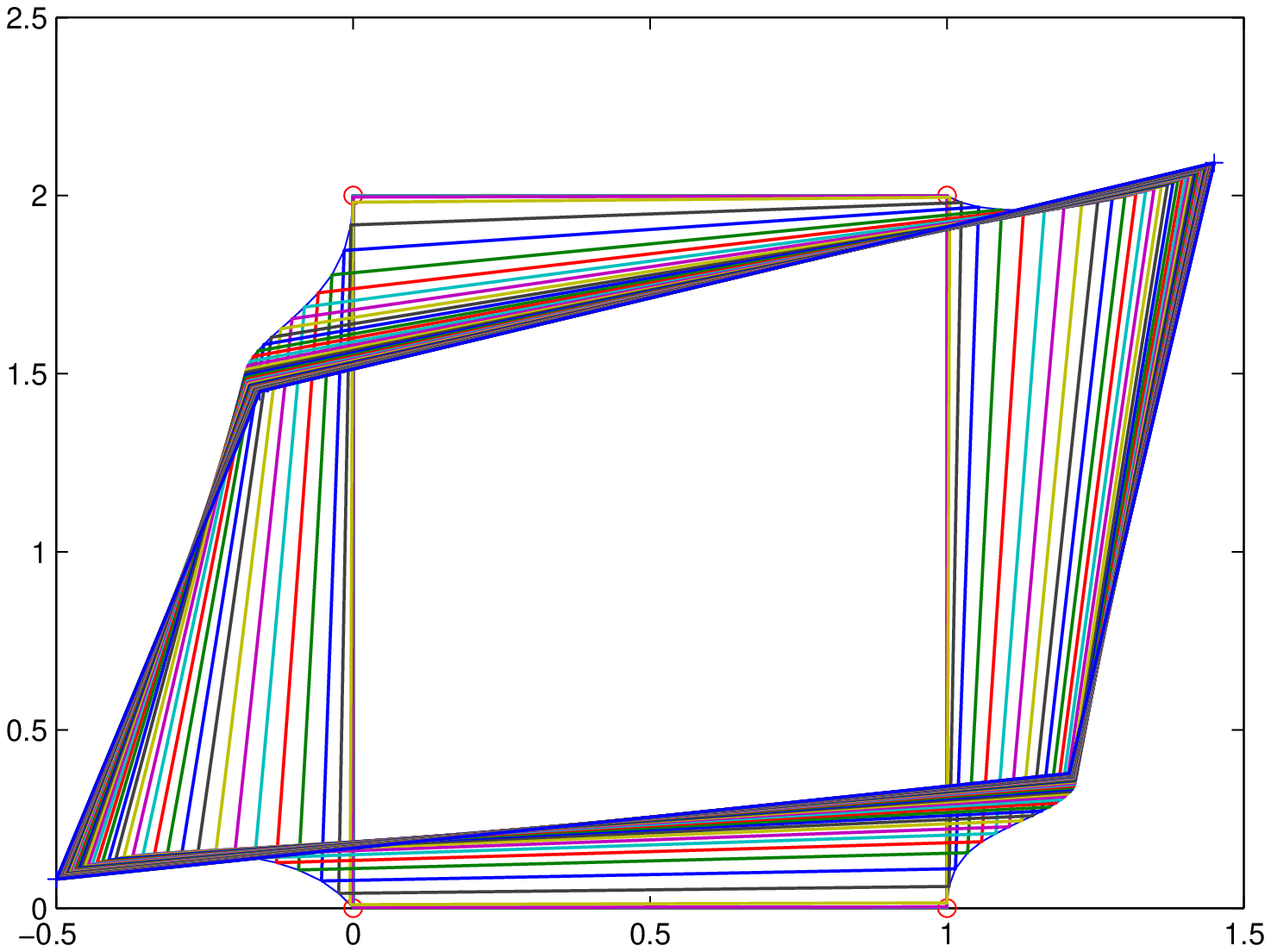}
\end{minipage}}
\subfigure[$s_c=0.5$]{\label{fig:invariant_s_09}
\begin{minipage}[b]{0.28\linewidth}
\centering
\includegraphics[width=2.8cm,height=2cm]{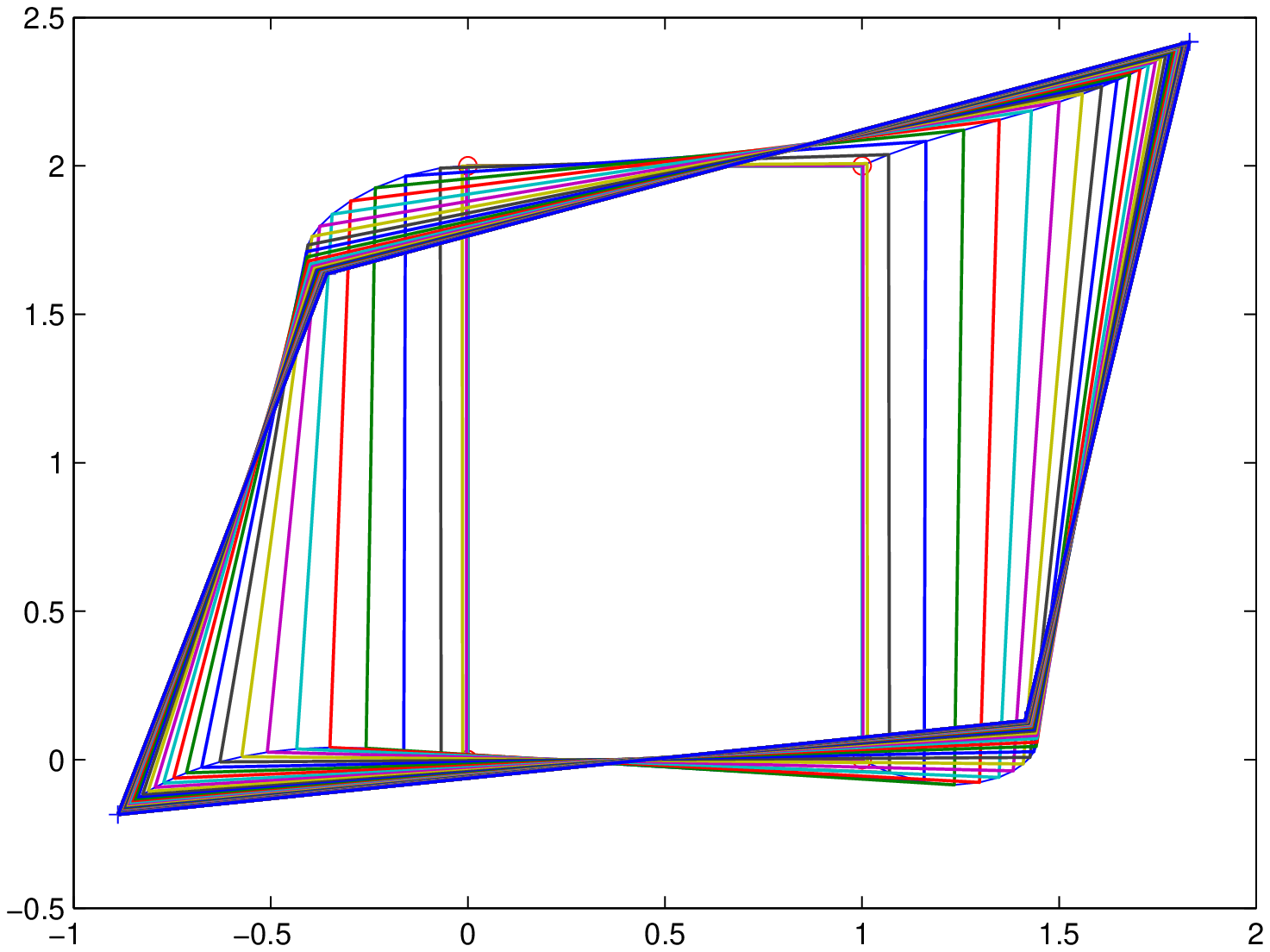}
\end{minipage}}
\caption{Formation system under time-invariant
scale}\label{fig:time_invariant}
\end{figure}

We show the energy saving of a formation system under the optimal
invariant scale $s_c^*$ and  compare it with other invariant scales.
Consider a formation system consists of four agents that are
initialized at $\mathbf{z}_0=[0;0;1;0;1;2;0;2]$ over a minimally
rigid graph $G_a$ in Fig. \ref{fig:four_topo_a}. The desired shape
corresponds to $G_a$ is measured by $\bar{S}=[3;3;3;2;3]$. The
time-invariant scales are scanned from 0.1 to 0.9 at step length 0.1
and the cost value $J_c$ at different $s_c$ are shown in Fig.
\ref{fig:time_invariant_s}. The data when $s_c=0.1, 0.3$ and 0.9 are
recorded in Table \ref{table:invar}. According to Fig.
\ref{fig:time_invariant_s}, the optimal time-invariant scale is
$s^*_c=0.5$ which is exactly the value calculated from
\eqref{eq:optStaticS}, and the optimal cost value is $J_c^*=2.1249$,
as shown in Table \ref{table:invar}. The formations when $s=0.5$ and
$s=0.9$ are shown in Fig. \ref{fig:invariant_s_05} and Fig.
\ref{fig:invariant_s_09} respectively. This verifies our conclusion
that \eqref{eq:optStaticS} helps to pick up the optimal geometry in
$\{\mathbf{e}|S\}$ and the control law \eqref{eq:myde_eqS} drives
the formation system attain this geometry when initialized within a
qualified neighborhood.

Moreover, we compare the differences between the cost functions
$J_c$ and $J_v$. The data in Table \ref{table:invar} infers that
when converge to the same geometry, formation system with
time-varying scale $\hat{s}^*(\be)$ has less cost value than system
with constant scale, which is measured by 1.7402 and 2.1249
respectively. This result verifies the advantages that are brought
by $\hat{s}^*(\be)$ in the nonlinear formation system.

We further observe the distance each agent travels during the
process and record the data in the table. According to the data, the
total distance the four agents travel has a relatively smaller value
of $1.8422$ when the scale is a time-varying function. On the other
hand, if we look at the maximal distance of the four agents travel
at different scales (as distinguished by underlines), agent 4 has
the smallest traveling distance of 0.6003 when $s_c\equiv0.5$.

%
%
\begin{table}
  \caption{Time-varying VS. Time-invariant scale under $G_a$}\label{table:invar}
  \begin{center}
\newsavebox{\tablebox}
\begin{lrbox}{\tablebox}
\begin{tabular}{|l|l|l|l|l|}
\hline
\multicolumn{5}{|c|}{Initial Conditions   $n=4; \mathbf{z}_0=[0;0;1;0;1;2;0;2]; S=[3;3;3;2;3]$} \\
\hline
 &$s_c$  & $J$  &  $L$: Length of each route & $Sum(L)$\\
\hline
\multirow{4}{*}{Invariant scale} &0.1 &5.5227 & [ 0.5167    0.7831    0.4686    $\underline{0.8318}$]&  2.6002\\
 &0.3 & 2.8944 & [0.4306  0.5534  0.3297  $\underline{0.6700}$] &   1.9837\\
 &0.9 & 5.1248 &[0.9559    0.6111    $\underline{0.9346}$    0.6331]& 3.1347\\
 &\textbf{\textbf{0.5 }} &  \textbf{2.1249 } &[0.5775  0.4608  0.4933  $\underline{0.6003}$] &2.1319\\
 \hline
Varying scale
 &\textbf{0.3}&\textbf{1.7402} & [0.3898 0.5190 0.2814 $\underline{0.6519}$] & 1.8422\\
\hline
\end{tabular}
\end{lrbox}
\scalebox{0.7}{\usebox{\tablebox}}
\end{center}
\end{table}


\subsection{The Topology, the initial geometry and the cooperative performance}
In the category of minimally rigid graphs, according to their
triangular complement graphs, they could be divided into different
subsets where graphs $G$ that belong to the same subset have the
same number of edges in $G^2$, i.e., $|E'|$ of $G'$ are identical.
In this subsection, parallel experiments are carried out over graphs
that are within one subset and between different subsets
respectively.

\begin{figure}
 \subfigure[Trajectories]{\label{fig:four_trajec}
\begin{minipage}[b]{0.45\linewidth}
\centering
\includegraphics[scale=.3]{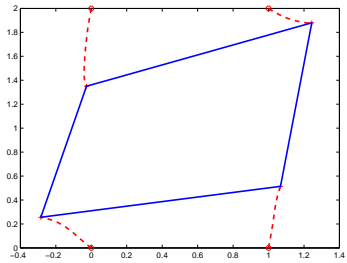}
\end{minipage}}
 \subfigure[Time-varying scale $\hat{s}^*(\be)$]{\label{fig:valueofS_four}
\begin{minipage}[b]{0.45\linewidth}
\centering
\includegraphics[scale=.3]{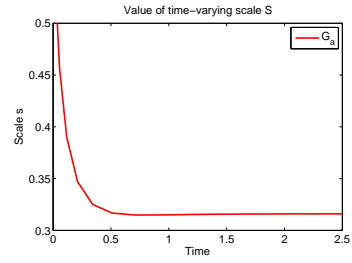}
\end{minipage}}
\caption{Four agents formations and the scale $\hat{s}^*(\be)$ over
$G_a$}\label{fig:four_formation}
\end{figure}

For comparison, the four agents in the previous experiment are
assigned with another minimally rigid graph shown in Fig.
\ref{fig:four_topo_b}, which also has one edge in its triangular
complement graph. The initial condition and the desired shape are
supposed to be the same as that of $G_a$. Note that for different
underlying graphs, the shape vector $S$ differs as well. When
applying the nonlinear control law \eqref{eq:myopt_multi} on systems
interconnected over $G_b$, the scale converges to 0.3563. The cost
value of the system over $G_b$ is 4.2120, compared to that of 1.7402
for $G_a$. The data is recorded in Table \ref{table:four}. Fig.
\ref{fig:four_formation} demonstrates the geometry of the agents
system and the trajectory of the scale $\hat{s}^*(\be)$. The circles
represent their initial positions
 and the bold blue lines are their final realization.

Apart from the cost value, Table \ref{table:four} also shows that
the total distance all agents travel is $1.8766$ for graph $G_b$,
which is larger than that of $G_a$.

Topology affects the system performance in terms of cost value and
the sum of lengths each node travel. According to the experimental
results, $G_a$ should be considered as being better than $G_b$ both
in cooperative performance and the total traveling distance.
Although the sub-optimization problem we discussed throughout the
paper depends highly on the initial conditions, the fact that $G_a$
outperforms $G_b$ stays true for all cases we tested with a large
number of different initial realizations.

\begin{figure}
 \subfigure[$G_a$, $|E'|=5$]{\label{fig:six_topo_a}
\begin{minipage}[b]{0.23\linewidth}
\centering
\includegraphics[scale=0.18]{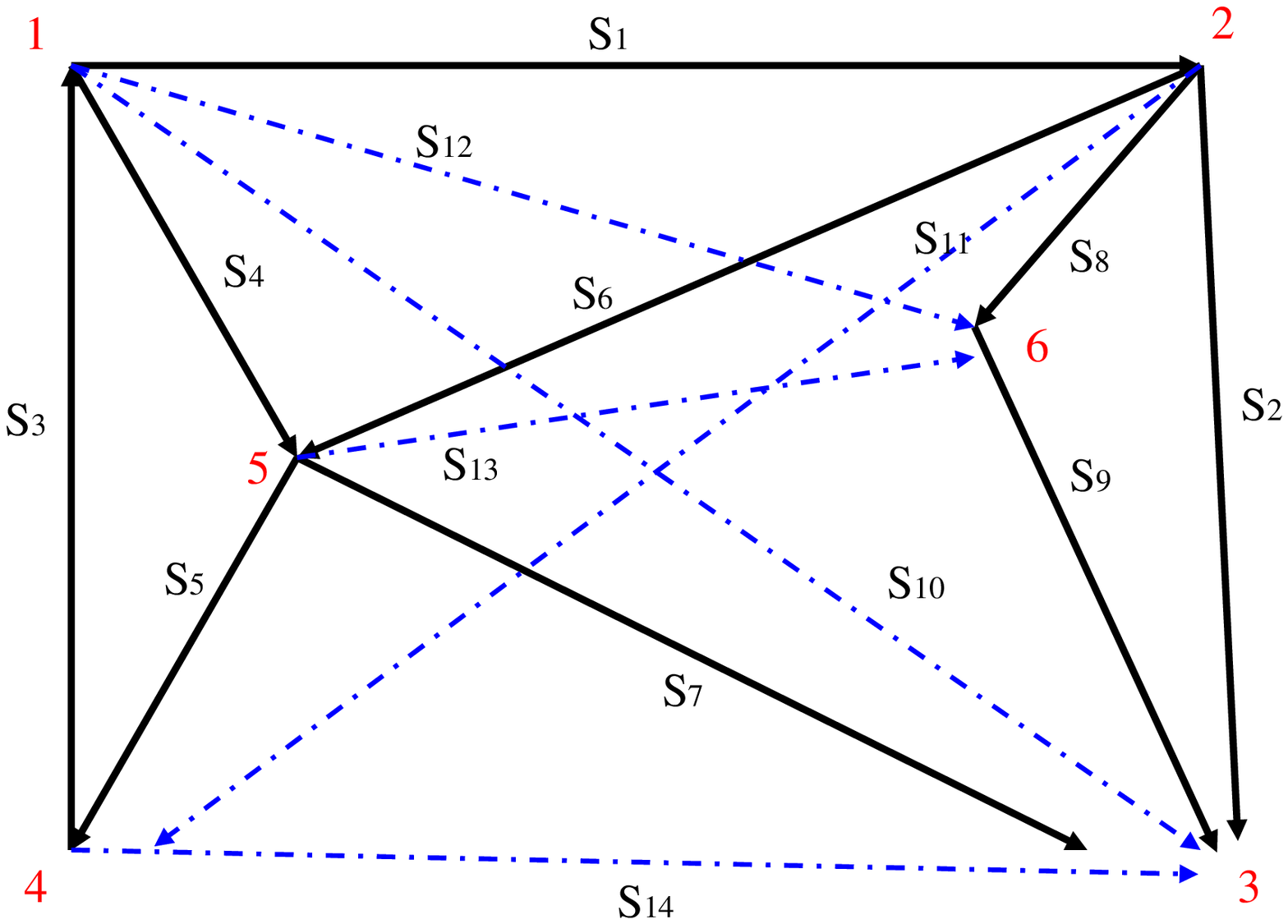}
\end{minipage}}
\subfigure[$G_b$, $|E'|=6$]{\label{fig:six_topo_b}
\begin{minipage}[b]{0.23\linewidth}
\centering
\includegraphics[scale=0.18]{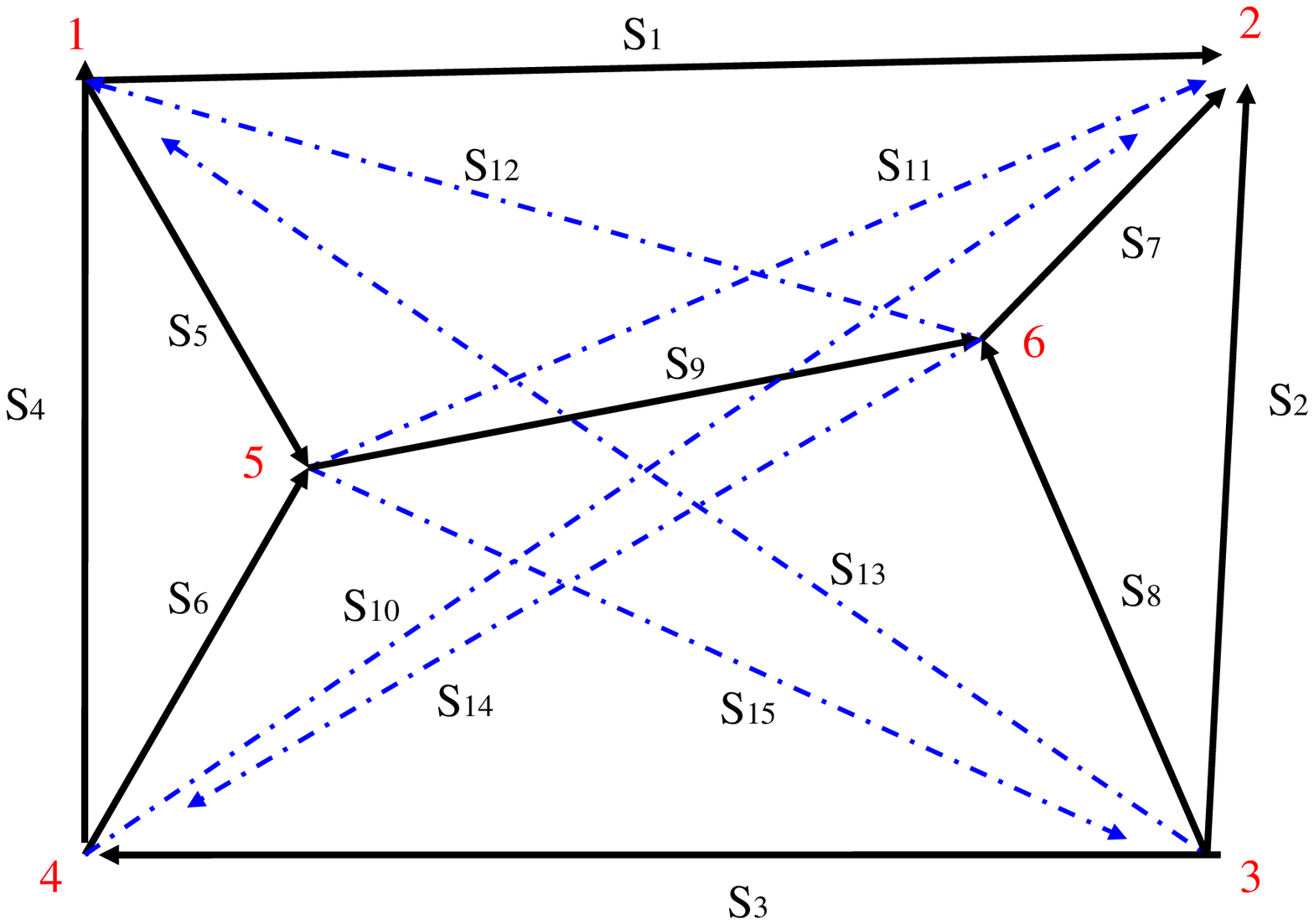}
\end{minipage}}
\subfigure[$G_c$, $|E'|=6$]{\label{fig:six_topo_c}
\begin{minipage}[b]{0.23\linewidth}
\centering
\includegraphics[scale=0.18]{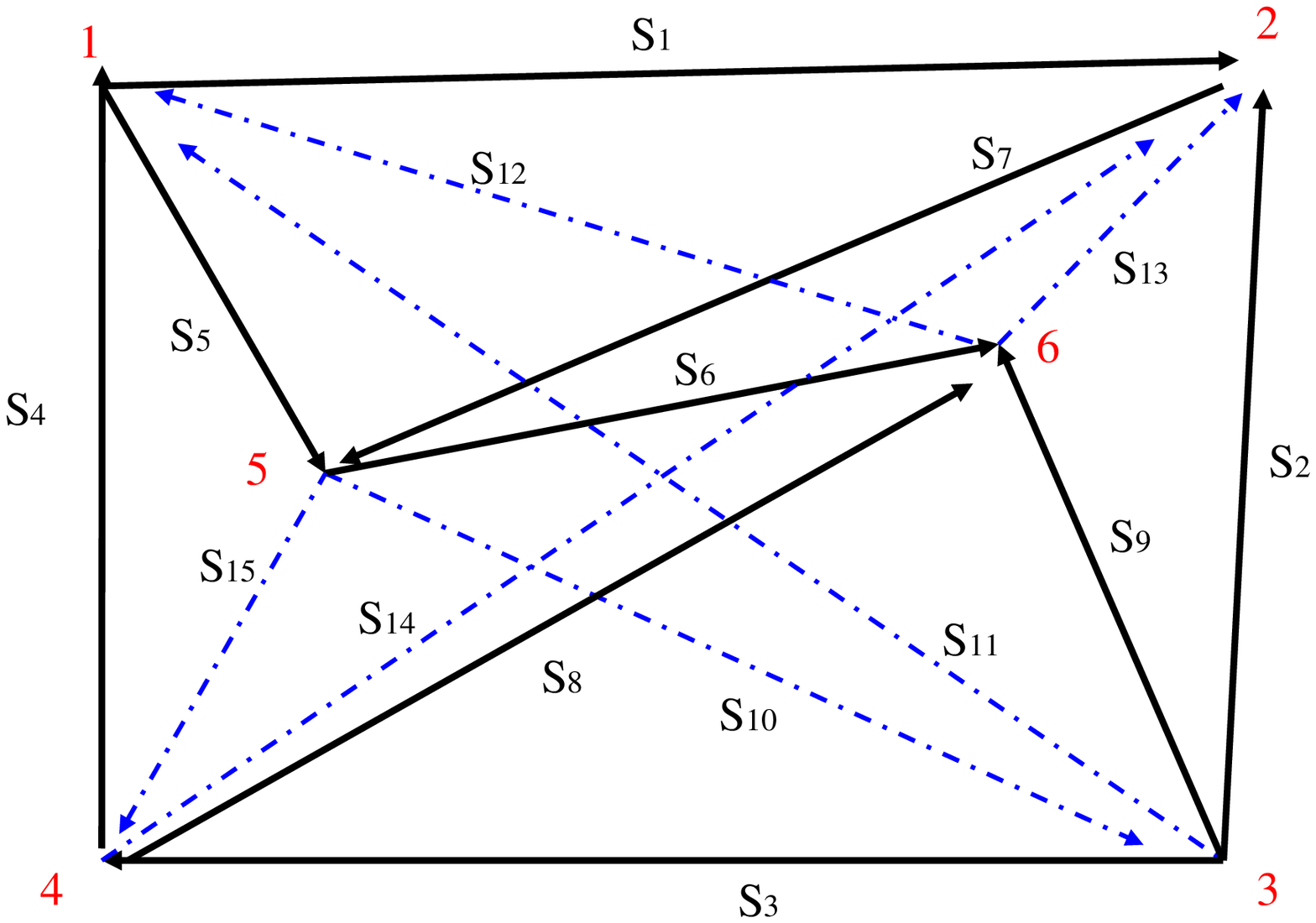}
\end{minipage}}
\subfigure[$G_d$, $|E'|=6$]{\label{fig:six_topo_d}
\begin{minipage}[b]{0.23\linewidth}
\centering
\includegraphics[scale=0.18]{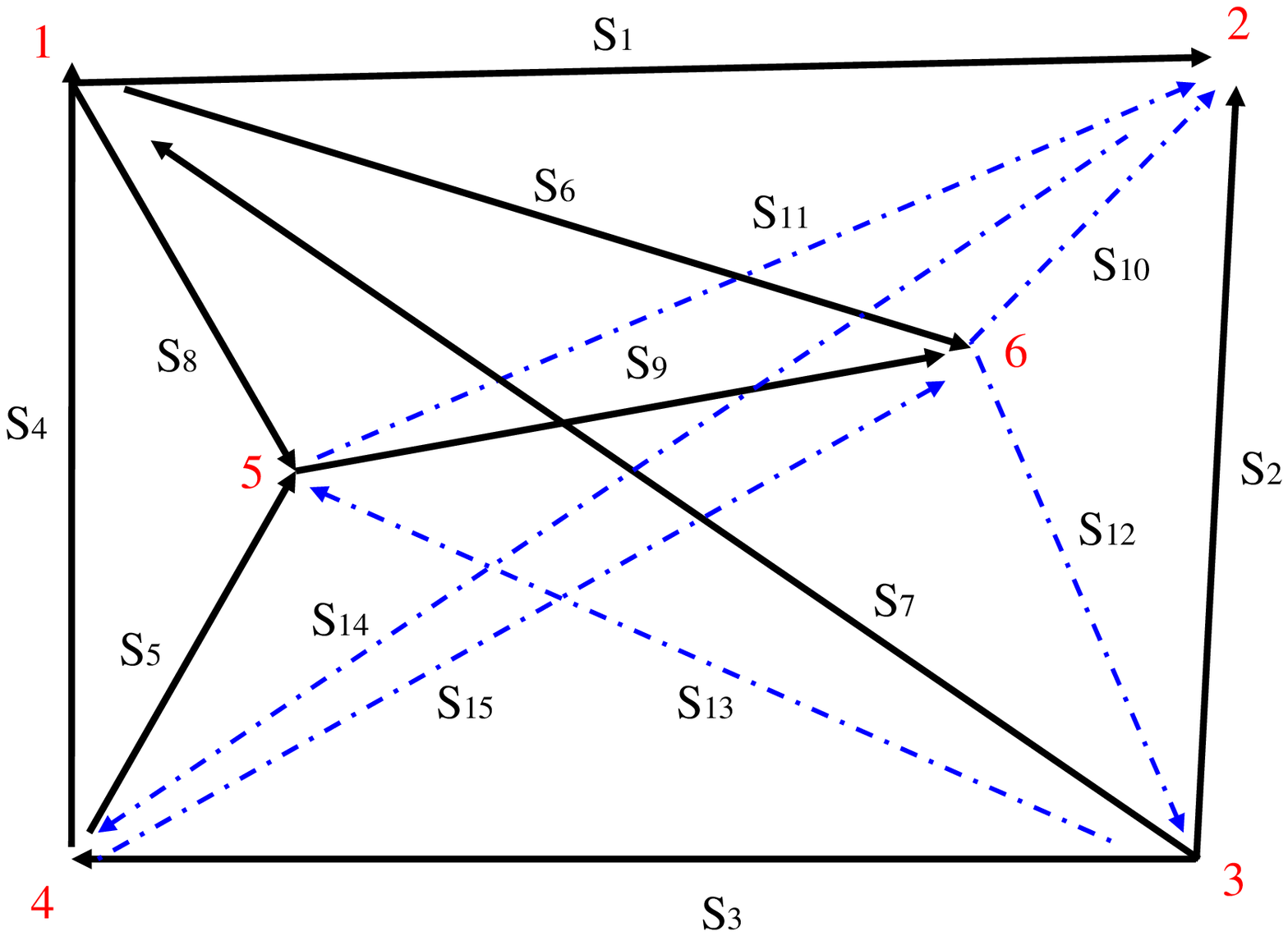}
\end{minipage}}
\caption{Four different graphs with different number of triangular
complement edges. The bold lines indicate the  minimally rigid
graphs, and dashed lines are their triangular complement
edges}\label{fig:six_topo}
\end{figure}

\begin{figure}
 \subfigure[$G_a$]{\label{fig:six_formation_a}
\begin{minipage}[b]{0.23\linewidth}
\centering
\includegraphics[scale=0.2]{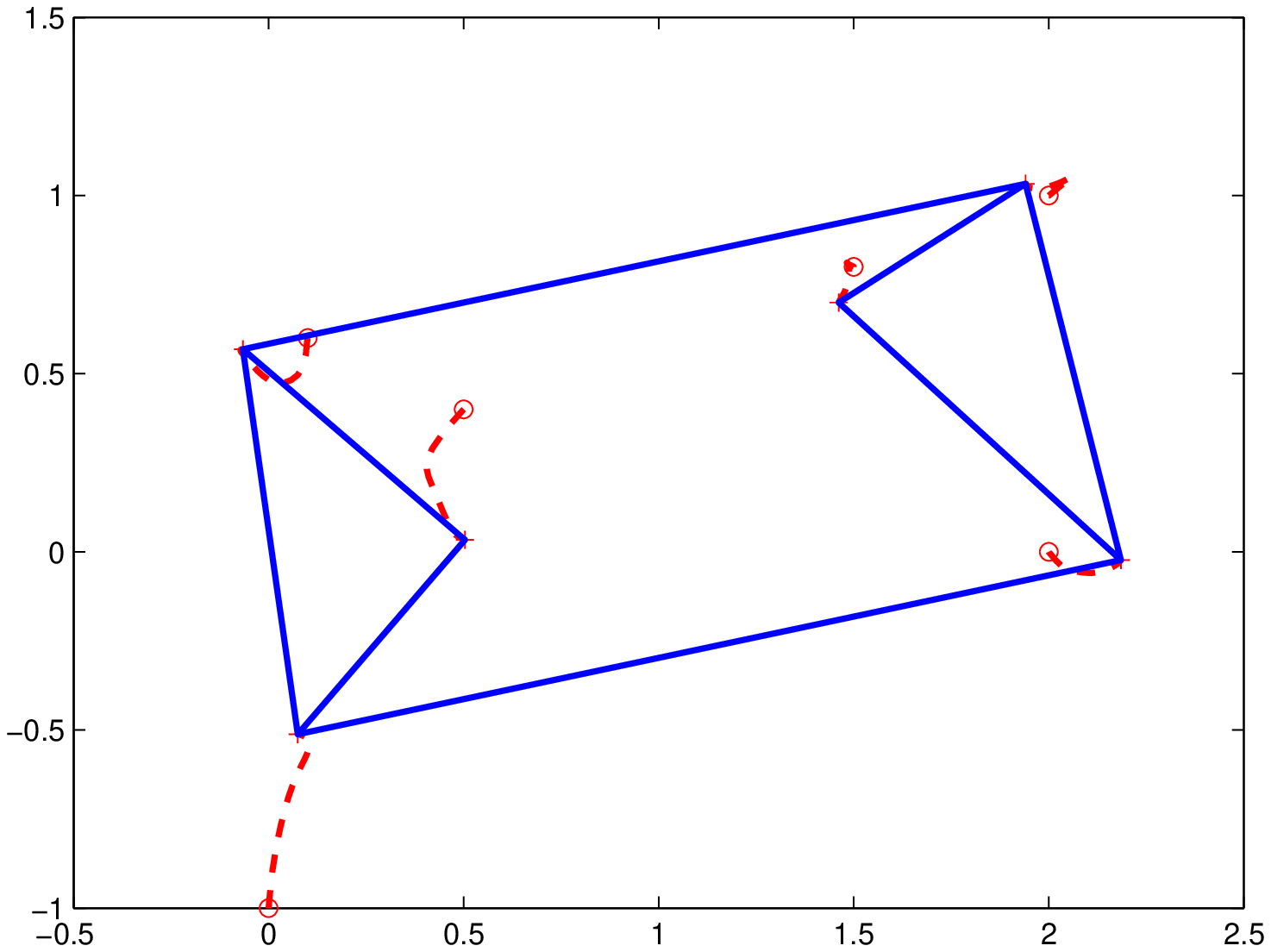}
\end{minipage}}
\subfigure[$G_b$]{\label{fig:six_formation_b}
\begin{minipage}[b]{0.23\linewidth}
\centering
\includegraphics[scale=0.2]{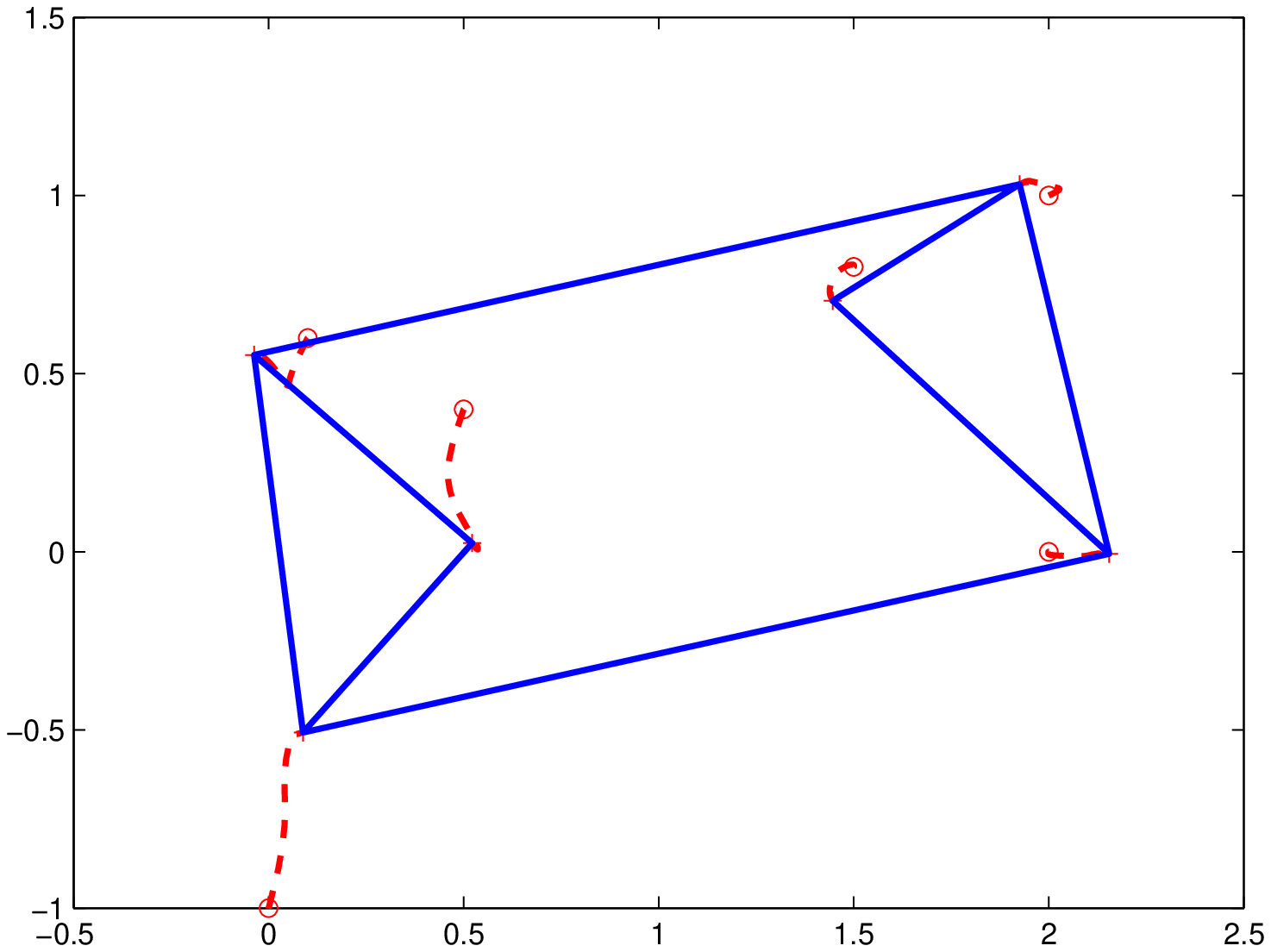}
\end{minipage}}
\subfigure[$G_c$]{\label{fig:six_formation_c}
\begin{minipage}[b]{0.23\linewidth}
\centering
\includegraphics[scale=0.2]{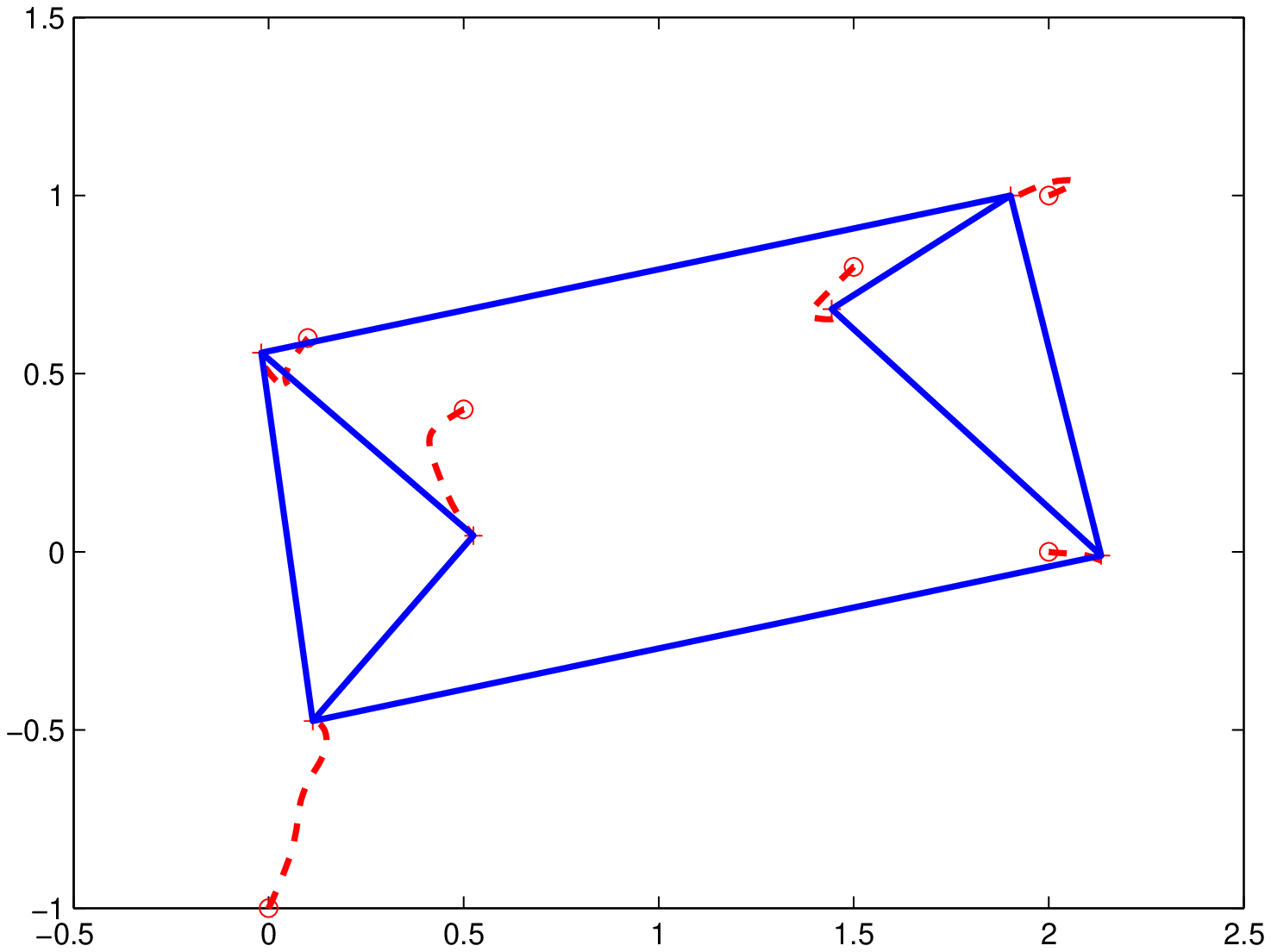}
\end{minipage}}
\subfigure[$G_d$]{\label{fig:six_formation_d}
\begin{minipage}[b]{0.23\linewidth}
\centering
\includegraphics[scale=0.2]{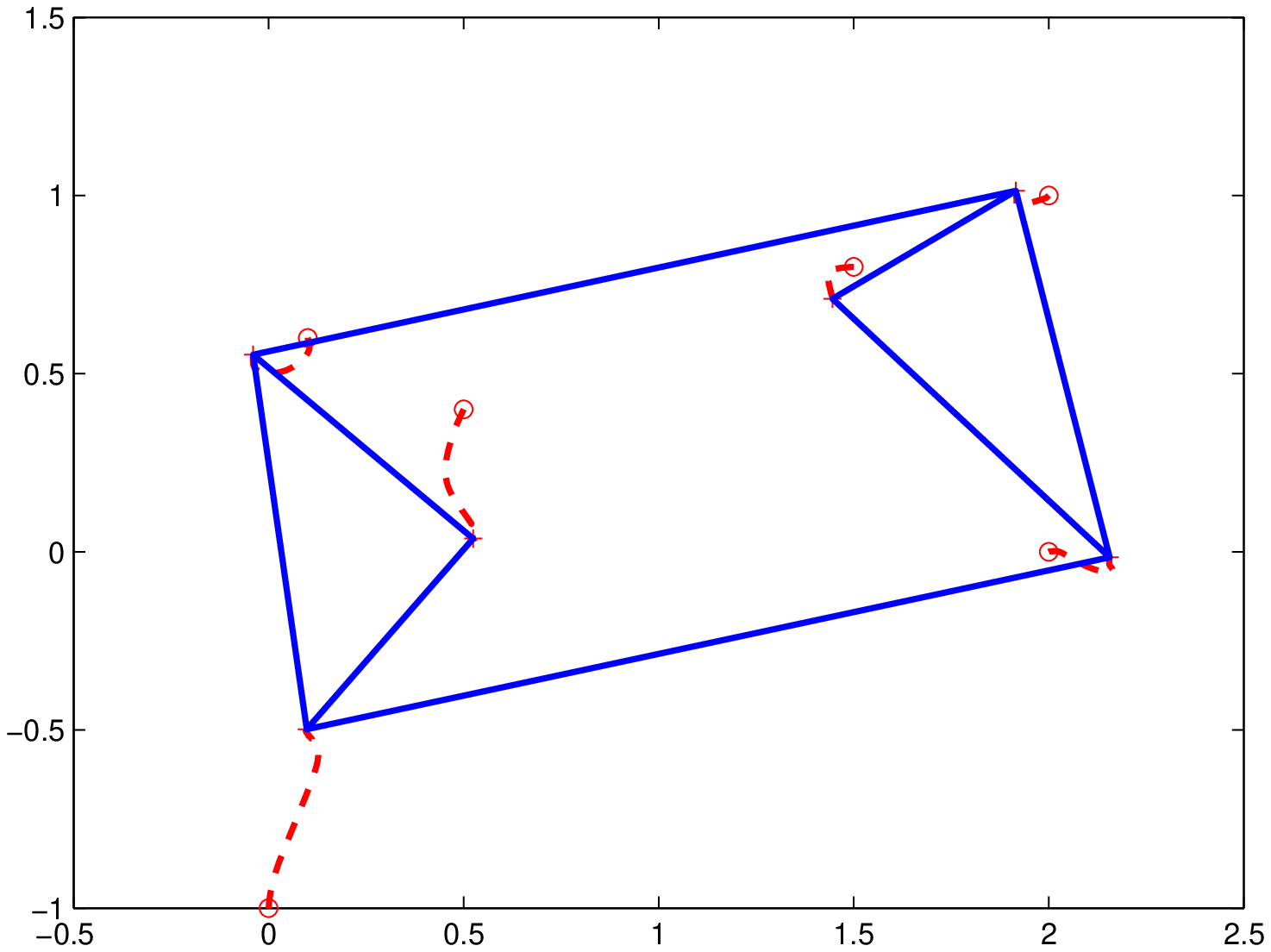}
\end{minipage}}
\caption{Trajectories of six agents interconnected over different
graphs}\label{fig:six_formation}
\end{figure}

\begin{figure}
\begin{center}
\includegraphics[scale=0.35]{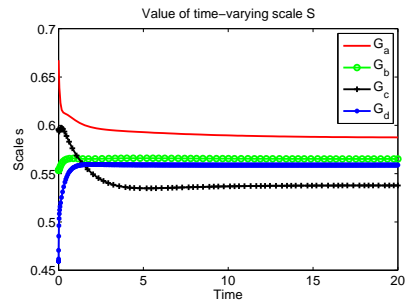}
\caption{Scale $\hat{s}^*(\be)$ over the four
graphs}\label{fig:valueofS}
\end{center}
\end{figure}

We further investigate another situation where  two underlying
graphs have different numbers of triangular complement edges. We
consider six agents over four different minimally rigid graphs, as
shown in Fig. \ref{fig:six_topo}.  The trajectories of the formation
system are shown in Fig. \ref{fig:six_formation} with experimental
results recorded in Table \ref{table:six}. The curves for the
time-varying scale $\hat{s}^*(\mathbf{e})$ over the four graphs are
demonstrated in Fig. \ref{fig:valueofS}.

A significant difference between $G_a$ and the other three graphs is
that $G_a$ only requires 5 additional edges for its triangular
complement graph while the number is 6 for the other three. This
interesting property explains the differences of he cost values and
total traveling distances in Table \ref{table:six}. Even with
different initial realizations $\mathbf{z}_0$ and $\mathbf{z}'_0$
that differs on the position of
 agent 4, the cost  value under $G_a$ is consistently smaller than that of the other three graphs, as
shown in the upper part and the bottom part of the table
respectively.

The data corresponding to the four graphs infer the possibility that
the system whose underlying graph having a smaller number of
triangular complement edges, such as $G_a$, may exhibit a better
cooperative performance than the one with more triangular complement
edges, such as $G_b$, $G_c$ and $G_d$.

On the other hand, the maximal distance the six agents travel, as
highlighted with underlines in the table, also proves the
superiority of graph $G_a$ over the others. The smallest maximal
distance agents travel is observed in $G_a$, which helps to prevent
the overwhelming of a single agent. This is again only a conjecture
from the current experiment without theoretical proofs, which is an
ongoing work to this research.

\begin{table}
  \caption{Systems over different underlying graphs with identical $|E'|$}
  \label{table:four}
\begin{center}
\begin{lrbox}{\tablebox}
\begin{tabular}{|l|l|l|l|l|}
\hline \multicolumn{5}{|c|}{$n=4; \mathbf{z}_0=[0;0;1;0;1;2;0;2];
S(G_a) =[3;3;3;2;3]; S(G_b)= [3;3;2; 4+\sqrt{15};3]$
} \\
\hline
& $\hat{s}^*_f$  & $J_v$  &  $L$: Length of each route & $Sum(L)$\\
\hline
$G_a$ &0.3159 & 1.7402&  [0.3898  0.5190  0.2814  $\underline{0.6519}$]  &  1.8422\\
$G_b$& 0.3563  &4.2120 & [0.3711  0.5013  0.3705  $\underline{0.6338}$]  &  1.8766\\
\hline
\end{tabular}
\end{lrbox}
\scalebox{0.7}{\usebox{\tablebox}}
\end{center}
\end{table}

\begin{table}
  \caption{Systems over different underlying graphs where $|E'|$ differs}
  \label{table:six}
 \begin{center}
\begin{lrbox}{\tablebox}
\begin{tabular}{|l|l|l|l|l|}
\hline \multicolumn{5}{|c|}{$n=6;
\bar{S}(G_a)=[3.6;1;1.01;0.52;0.41;2.61;2.41;0.29;0.89];$}\\
\multicolumn{5}{|c|}{$\bar{S}(G_b)
=[3.6;1;4;1.01;0.52;0.41;0.29;0.89;1.16]$} \\
\multicolumn{5}{|c|}{
$\bar{S}(G_c)=[3.6;1;4;1.01;0.52;1.16;2.61;2.89;0.89];$}\\
\multicolumn{5}{|c|}{$\bar{S}(G_d)
=[3.6;1;4;1.01;0.41;2;4.6;0.52;1.16]$} \\
\hline
\multicolumn{5}{|c|}{$\mathbf{z}_0=[0.1;0.6;2;1;2;0;0;-1;0.5;0.4;1.5;0.8]$;}\\
\hline
& $\hat{s}^*_f$  & $J_v$  &  $L$: Length of each route & $Sum(L)$\\
\hline
$G_a$ &0.5873 & \textbf{0.6918} & [0.3143  0.2128  0.2214  $\underline{0.5202}$  0.4290  0.1411]& 1.8388\\
$G_b$& 0.5652 & 0.7724  &[0.2805  0.1532  0.1737  $\underline{0.5289}$  0.4500  0.1488] &1.7351\\
$G_c$ & 0.5369& 0.9152&  [0.2900  0.2767  0.1799  $\underline{0.5628}$  0.4139  0.2713]  &  1.9946\\
$G_d$& 0.5589  &0.9169 & [0.2407  0.1185  0.2152  $\underline{0.5305}$  0.3916  0.1189]  &  1.6154\\
\hline \multicolumn{5}{|c|}{$\mathbf{z}'_0=[0.1;0.6;2;1;2;0;-1;0;0.5;0.4;1.5;0.8]$;}\\
\hline
& $\hat{s}^*_f$  & $J_v$  &  $L$: Length of each route & $Sum(L)$\\
\hline
$G_a$ &0.6105 & \textbf{1.0284}  &[0.5100  0.2428  0.1906  $\underline{0.7621}$  0.3985 0.1467]& 2.2507\\
$G_b$& 0.5838  &1.5533  &[0.6338  0.4744  0.5584  $\underline{0.7953}$  0.3977 0.3489] &3.2086\\
$G_c$& 0.5421  &1.7603 & [0.4980  0.5902  0.4058  $\underline{0.8065}$  0.4816 0.4097] &3.1918\\
$G_d$& 0.5570  &2.4951&  [0.5947  0.2069  0.4649  $\underline{0.8142}$  0.3718 0.4263] &2.8788\\
\hline
\end{tabular}
\end{lrbox}
\scalebox{0.7}{\usebox{\tablebox}}
\end{center}
\end{table}

\begin{figure}
\begin{center}
\includegraphics[scale=.35]{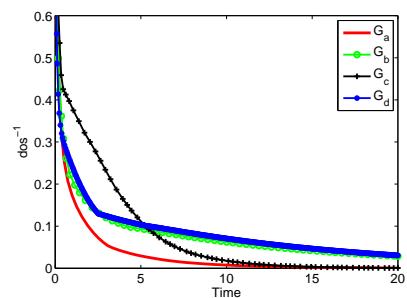}
\caption{The inverse of the geometry's DOS  at time $t$}
\label{fig:for_bias_six}
\end{center}
\end{figure}

The DOSs of the four graphs $G_a$, $G_b$, $G_c$ and
$G_d$ to the desired shape  are shown in Fig. \ref{fig:for_bias_six}. The solid line,
which corresponds to graph $G_a$, has both a fast convergence speed
and stay in a high DOS during the process. Although system over
$G_c$ achieved the desired shape as fast as the one over $G_a$ did,
its geometric distance to the desired shape during convergence was much larger.

\subsection{Geometries in sensor-target localization}
A triangle that has a higher DOS to the equilateral
one could provide better estimation of the target location, as
discussed in Section \ref{sec:localization}.

\begin{figure}
 \subfigure[The inverse of  the triangles DOS  to the equilateral one]{\label{fig:localization_fe}
\begin{minipage}[b]{0.45\linewidth}
\centering
\includegraphics[scale=.3]{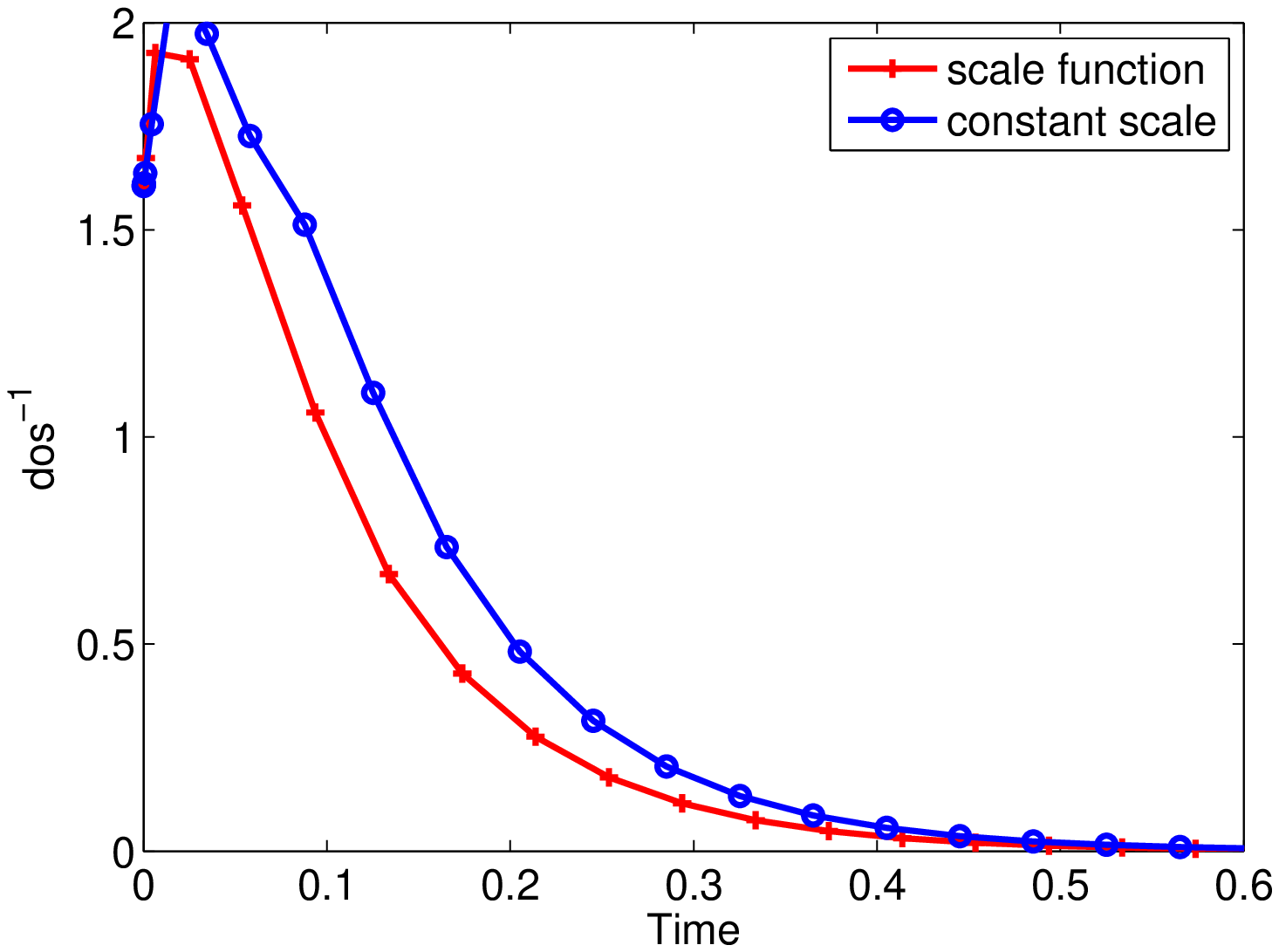}
\end{minipage}}
 \subfigure[Determinant of the Fisher information matrix]{\label{fig:localization_fish}
\begin{minipage}[b]{0.45\linewidth}
\centering
\includegraphics[scale=.3]{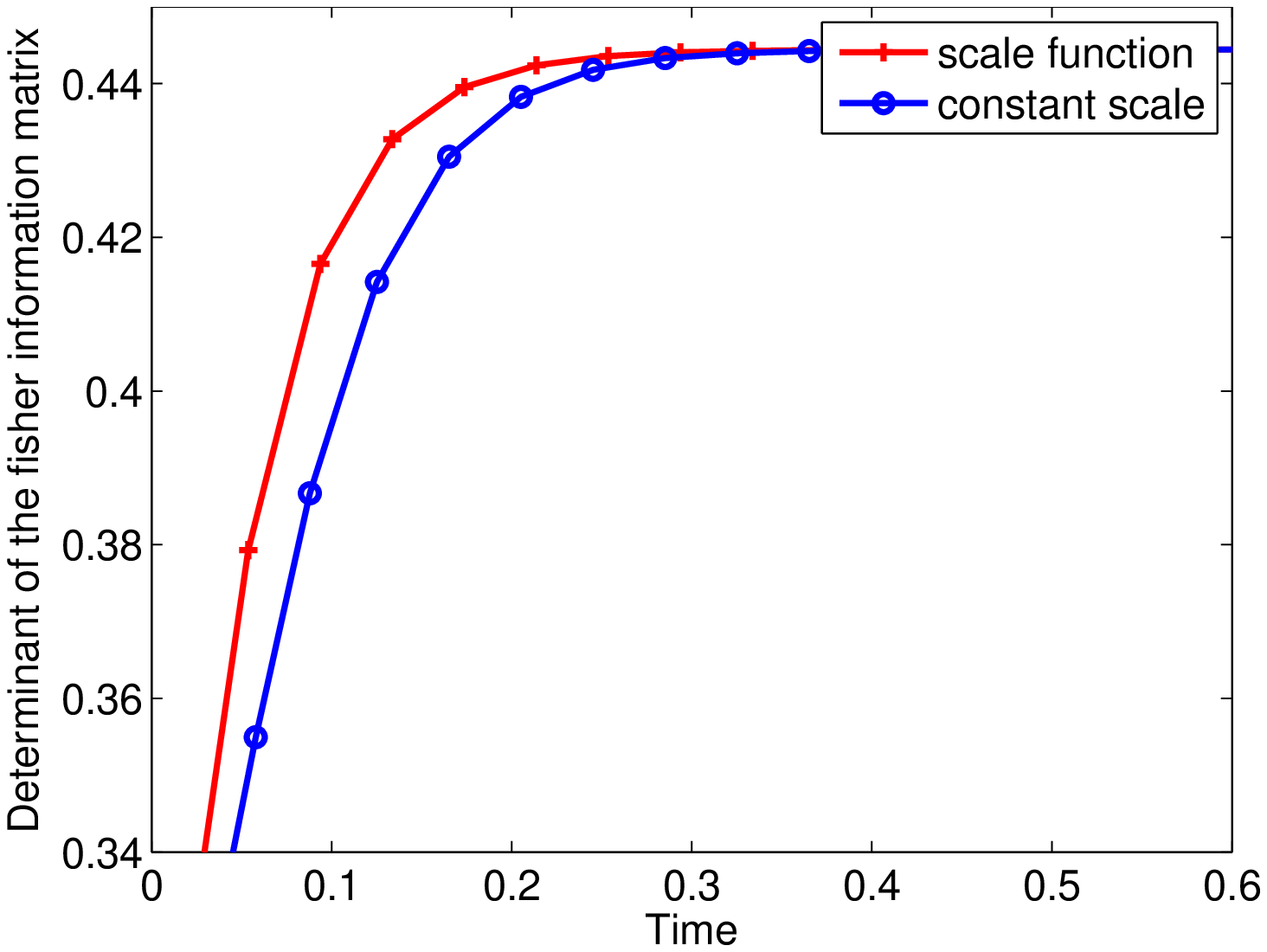}
\end{minipage}}
\caption{Formation control in bearing-only sensor-target
localization}
\end{figure}

We consider three agents initialized at
$\mathbf{z}_0=[0;0;3;1.5;4;0]$. The desired shape is a equilateral
triangle, which is one of the optimal shapes in bearing-only
sensor-target localization. The specialty of equilateral triangle  For systems with the time-varying scale
function and with a constant scale respectively, the geometries'
DOS during the process are shown in Fig.
\ref{fig:localization_fe}.

The determinant of the Fisher information matrix during the process
is calculated in Fig. \ref{fig:localization_fish}. The system with a
constant scale has consistently smaller degree of
similarity to the equilateral triangle, and also smaller determinant
value, which indicates the variance estimation is relatively worse.
This validates the conclusion of Corollary \ref{cor:similarity}.

\section{Conclusions and Future Works}
For a formation system with a flexible scale but under shape
constraints, we discussed the strategies of choosing the desired
geometries' scale so as to achieve better cooperative performance.
In order to ensure the exponential stability of the system over a minimally rigid graph,
fixed-structured nonlinear control laws on the edges and the scale were considered, where two
types of scale design methods were proposed, namely  the
time-invariant scale and the time-varying scale function.  It was
proved that a system with the time-varying scale function could
further reduce the minimum of the cost value when the scale is
constant. By defining a triangular complement graph, the algorithm
also applied to the multiple agents case. The controllability of the
formation system was also discussed by adjusting the convergence
rate of each edge with respect to the reference system.

These results were validated on various simulations where we
compared the cost values of  systems with constant scale and the
time-varying scale function respectively.  The experimental results
also inferred the possibility that the underlying topology with a
smaller number of triangular complement edges may always cost less
during convergence and thus may have better cooperative performance.
Theoretical proof to this conjecture is still undergoing. Moreover,
we also applied the control laws in sensor-target localization to
show the prominent features of the nonlinear formation system we
designed.

The nonlinear dynamics \eqref{eq:myopt_multi} is a centralized
control law as $\hat{s}^*(\mathbf{e})$ contains states information
of every agent in the system rather than local information.
Decentralized/distributed algorithms should be considered in the
future so as to meet the scalability requirements. Moreover, further
research should focus on eliminating the restriction on the
minimally rigid underlying graph.

\ack The authors would like to thank Professor Brian Anderson for the insightful comments and discussions, which have significantly improved the presentation of this work.

\bibliographystyle{wileyj}
\bibliography{E:/papers/odds/odds}

%

\end{document}